\documentclass{article}
\usepackage{url}
\usepackage{hyperref}
\usepackage{booktabs}
\usepackage{float}
\usepackage{amsmath}
\usepackage{ascmac}
\usepackage{amssymb}
\usepackage[nameinlink]{cleveref}
\usepackage{braket}
\usepackage{physics}
\usepackage{mathrsfs}
\usepackage{cite}
\usepackage{tikz}
\usetikzlibrary{arrows.meta, positioning}
\usepackage{graphicx}
\usepackage{bm}
\usepackage{color}
\usepackage{amsfonts}
\usepackage{wrapfig}
\usepackage{subcaption}
\usepackage{comment}
\usepackage{authblk}

\usepackage{mathtools}
\numberwithin{equation}{section}
\usepackage[english]{babel}
\usepackage{geometry}
\newcommand{\brho}{\bar{\rho}}
\newcommand{\mA}{\mathcal{A}}
\newcommand{\mQ}{\mathcal{Q}}
\newcommand{\hn}{\hat{n}}
\newcommand{\erfc}{\mathrm{erfc}}

\newcommand{\address}[2][]{\affil[#1]{#2}}
\title{\Large\bfseries Non-stationary current fluctuations in 1D boundary-driven diffusive systems via Macroscopic Fluctuation Theory}

\author[1]{Daisuke Suzuki\footnote{suzuki.d.2e67@m.isct.ac.jp}}
\author[1]{Tomohiro Sasamoto\footnote{sasamoto.t.0c12@m.isct.ac.jp}}
\address[1]{Department of Physics, Institute of Science Tokyo, 2-12-1 Ookayama, Meguro-ku, Tokyo 152-8551, Japan}
\date{}
\begin{document}
\maketitle
\vspace{-20pt}
\begin{abstract}
 While Macroscopic Fluctuation Theory (MFT) has been highly successful in analyzing non-equilibrium steady states, its application to non-steady-state processes remains limited. In this study, we apply MFT to the relaxation process of one-dimensional boundary-driven diffusive systems coupled to particle reservoirs at both ends. We exactly derive the current variance for systems with a constant diffusion coefficient and arbitrary mobility, as well as the cumulant generating function for the current in Reflective Brownian Motion (RBM). Our results demonstrate that non-steady current fluctuations during the approach to a steady state can be quantitatively described within the MFT framework.
\end{abstract}
\tableofcontents

\section{Introduction}

One of the primary goals of non-equilibrium statistical mechanics is to construct a unified framework for describing the fluctuations of physical quantities, analogous to the Boltzmann principle in equilibrium systems. In non-equilibrium systems, the currents of particles or energy play an essential role that is absent in equilibrium systems; thus, elucidating their statistical properties is a critical step toward extending the understanding of non-equilibrium systems. In particular, the calculation of the large deviation function is a central challenge. This is because it governs the system's statistical behavior and acts as a "non-equilibrium thermodynamic potential"\cite{Bodineau2007,Derrida2007,Bertini2009,Jona-Lasinio2010, Derrida2025}. 

To date, one-dimensional diffusive systems such as the Symmetric Simple Exclusion Process (SEP)\cite{Spitzer1970} and Reflective Brownian Motion (RBM)\cite{Harris1965} have been extensively studied as minimal models. These studies have contributed significantly to the exploration of universal properties of non-equilibrium fluctuations, including the additivity principle in finite systems driven into a non-equilibrium steady state (NESS) \cite{Bodineau2004,Bertini2006}, and the Gallavotti-Cohen type symmetry appearing in the current scaled cumulant generating functions (SCGF) of NESS and certain types of non-stationary (semi-)infinite systems\cite{gallavotti_dynamical_1995,Lebowitz1999,Derrida2009a,Saha2023}.

However, large deviations in finite non-stationary systems remain poorly understood. This is a common situation in nature, where both ends of a system are connected to reservoirs. In these finite systems, the contributions from the left and right reservoirs interfere with each other. In the regime $T\ll L^2$, the interference between the two reservoirs is negligible, and $Q_T$ grows as $O(\sqrt{T})$. However, for $T\gg L^2$, the system has nearly reached a non-equilibrium steady state, and $Q_T$ grows as $O(T)$. Namely, as the system relaxes toward a steady state, the scale of the integrated current during time $[0,T]$ changes from $O(\sqrt{T})$ to $O(T)$. In this process, the system gradually loses the memory of its initial state, and the boundary driving begins to dominate the dynamics. Analyzing current fluctuations in this crossover region, where initial and boundary contributions are coupled, helps us understand how boundaries and initial conditions affect the formation of NESS. Therefore, studying current large deviations during relaxation provides new insights into non-equilibrium fluctuations.

\begin{figure}[H]
  \centering
  \begin{minipage}[c]{0.55\textwidth}
    \centering
    \includegraphics[width=\textwidth]{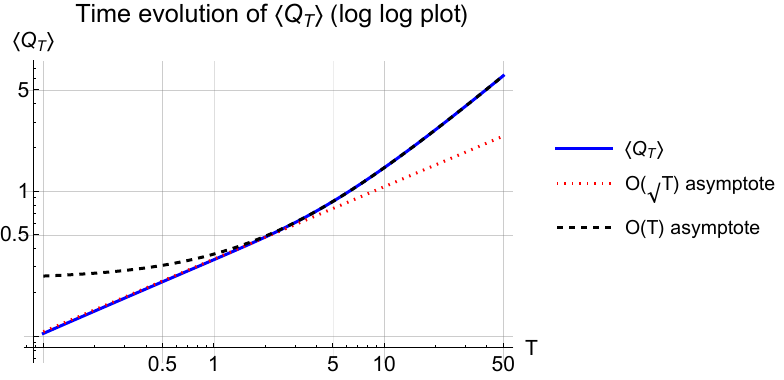}
  \end{minipage}
  \hfill 
  \begin{minipage}[c]{0.4\textwidth}
    \caption{An example of the time evolution of the expected integrated current $\ev{Q_T}$ in boundary-driven RBM. $Q_T$ crosses over from $O(\sqrt{T})$ growth to linear $O(T)$ growth in time.}
    \label{fig:current_t_dependence}
    \vspace{10pt}
  \end{minipage}
\end{figure}

Macroscopic Fluctuation Theory (MFT), recently proposed by Bertini et al., has emerged as a powerful theoretical framework for analyzing large deviations of density and current fields in these diffusive systems\cite{Bertini2002,Bertini2015}. MFT describes fluctuations in macroscopic systems using the Martin-Siggia-Rose path integral formalism\cite{Martin1973}, mapping the fluctuations to a classical action. This approach allows the tilted trajectory to be obtained by minimizing the action, thereby reducing the problem of determining statistical properties to solving a set of coupled Hamilton's equations with certain boundary conditions (The detail is explained in section 2). Despite its utility, MFT has primarily been applied to non-equilibrium steady states\cite{Bertini2002,Bertini2005,Bertini2006,Derrida2011,carinci_large_2023,capanna_class_2024}, and its application to non-stationary systems remains limited. While there are successful derivations of current CGF for infinite\cite{Mallick2022} and semi-infinite SEP\cite{Grabsch2024,sharma_large_2026}, there are currently no examples of its application to finite non-stationary systems. Extending MFT to such systems is essential for advancing our understanding of the statistical mechanics of relaxation dynamics. Moreover, although the quantitative agreement between MFT and microscopic calculations in NESS and infinite non-stationary systems has provided a strong foundation for the theory , its applicability to finite non-stationary regimes—where the interplay between initial conditions and boundary driving is most prominent—remains to be fully explored. Quantitative verification in this challenging regime is essential for establishing MFT as a robust and universal framework for describing non-equilibrium relaxation processes. \\

In this paper, we demonstrate that non-stationary current fluctuations during the relaxation process can be systematically analyzed within the MFT framework. In Section 2, we formulate the MFT approach for finite non-stationary systems and derive the corresponding MFT equations and boundary conditions. In Section 3, we derive a general expression for the current variance in diffusive systems with a constant diffusion coefficient and arbitrary mobility. We consider two different initial conditions: annealed initial condition and quenched initial condition. In Section 4, we provide an exact derivation of the scaled cumulant generating function (SCGF) for the current in RBM, where the MFT equations are analytically solvable. We also discuss coupling strength with the reservoir and initial condition.

\section{Macroscopic Fluctuation Theory}\label{section2}

\subsection{Overview of MFT }\label{sec; MFT}

Macroscopic Fluctuation Theory (MFT) is a framework to analyze large deviations of diffusive many-body systems\cite{Bertini2002,Bertini2015}. In this section, we outline the procedure for calculating the SCGF of the integrated current $Q_T$ within the framework of MFT. Here, we introduce two transport coefficients: the diffusion coefficient $D(\rho)$ and the mobility $\sigma(\rho)$. Let us consider a system of length $L$ coupled to two particle reservoirs with densities $\rho_L$ and $\rho_R$ at its boundaries. The coefficients $D(\rho)$ and $\sigma(\rho)$ are defined in terms of the expectation and variance of the integrated current $Q_T$ over time $T$ as follows\cite{Bodineau2004,Derrida2007}:
\begin{align}
  \lim\limits_{T\rightarrow\infty}\frac{\ev{Q_T}}{T}&=\frac{D(\rho)}{L}\Delta\rho\qquad \text{for}\ \rho_R-\rho_L\ \text{small} \label{eq:Drho}\\
  \lim\limits_{T\rightarrow\infty}\frac{\ev{Q_T^2}}{T}&=\frac{\sigma(\rho)}{L}\qquad \text{for} \ \rho_L=\rho_R \label{eq:sigmarho}
\end{align}
Macroscopically, the diffusive system is governed by the following diffusion equation\cite{Bertini2015}:
\begin{align}\label{eq;diffusioneq}
  \partial_t\rho=\partial_x\qty(D(\rho)\partial_x\rho-\frac{\sigma(\rho)}{2}E(t)),
\end{align}
where, $E(t)$ is the external field. MFT provides a way to evaluate large fluctuations from the diffusion equation by the large deviation principle. In the present work, we restrict our attention to the case without an external field $(E(t)=0)$.

The diffusion equation describes the macroscopic time evolution of the expected density in non-equilibrium many-body systems that exhibit diffusive behavior. In other words, it can be viewed as the coarse-grained, large-system limit of a microscopic model of interacting particles, where fluctuations are neglected through a suitable scaling. In this section, we adopt stochastic lattice gases as microscopic models for such systems. To derive macroscopic behavior from a discrete lattice model, a specific scaling procedure is required.
To extract the macroscopic behavior by coarse-graining the microscopic model, we introduce a diffusive scaling that relates the microscopic coordinates $(i,\tau)$ $(0<i<L,\ 0<\tau<T)$ to the macroscopic coordinate $(x,t)$ $(0<x<L,\ 0<t<T)$ using a large scaling parameter $\Lambda$ as follows\cite{Spohn1991,Kipnis1999}:
\begin{align}\label{eq;diffusivescale}
  x=\frac{i}{\Lambda},\ t=\frac{\tau}{\Lambda^2} \quad \text{for} \quad 0<i<\Lambda L, \ 0<\tau<\Lambda^2T.
\end{align}
It also corresponds to take large system limit keeping with $t/x^2=\text{const}$.
By taking the limit $\Lambda\rightarrow\infty$ (hydrodynamic limit), the discrete microscopic dynamics converge to a continuous macroscopic description: diffusion equation $\eqref{eq;diffusioneq}$.

Let $Q_T$ be the net number of particles passing through a certain point during the time interval $[0,T]$. In the diffusive scaling limit $\Lambda\rightarrow\infty$ defined above, $Q_T$ satisfies the large deviation principle $P(Q_T/\Lambda=q)\asymp e^{-\Lambda I(q)}$ where $I(q)$ is the large deviation function of $Q_T$. This function $I(q)$ is related to the SCGF, defined as $\mu(\lambda)\equiv\lim\limits_{\Lambda\rightarrow\infty}\frac{1}{\Lambda}\ln\ev{e^{\lambda Q_T}}$, via the Legendre transformation $I(q)=\sup\limits_{\lambda}[q\lambda-\mu(\lambda)]$ \cite{Touchette2009}. Consequently, in the limit $\Lambda\rightarrow\infty$, the following relationship holds:
\begin{align}\label{eq;dmu}
  \dv{\mu}{\lambda}=q\simeq\frac{Q_T}{\Lambda}\quad (\Lambda\rightarrow\infty).
\end{align}
Here, $q$ is the scaled integrated current. To derive the current SCGF $\mu(\lambda)$ within the MFT framework, we express the moment-generating function for current $\ev{e^{\lambda Q_T}}$ in terms of a path-integral representation by introducing the auxiliary field $H(x,t)$ \cite{Krapivsky2014}, specifically leveraging the Martin-Siggia-Rose (MSR) formalism\cite{Martin1973}
\begin{align}
  \ev{e^{\lambda Q_T}}=\int\mathcal{D}[\rho, H]e^{-\Lambda S[\rho,H]}.
\end{align}
The functional $S[\rho,H]$ is known as the MFT action. By considering $\Lambda\rightarrow\infty$ limit and from saddle point analysis, 
\begin{align}
  \int\mathcal{D}[\rho, H]e^{-\Lambda S[\rho,H]}\simeq e^{-\Lambda S[\rho,H]}
\end{align}
In the above equation, $\rho$ and $H$ on the left-hand side denote independent fields, whereas those on the right-hand side represent the optimal trajectories satisfying the variational equation $\delta S[\rho,H]=0$. From the variational equation $\delta S[\rho, H]=0$, a set of two coupled non-linear partial differential equations (MFT equations) and boundary conditions are obtained (see Section 2.2 for details). 
The solution to the MFT equations represents the optimal trajectory of $\rho(x,t)$ and $H(x,t)$. From the MFT equations, it is shown that the tilted current $j$ can be expressed as $j=-D(\rho)\partial_x\rho+\sigma(\rho)\partial_xH$. Here, $\partial_x H$ represents an effective external field that generates the tilted dynamics. Since the scaled integrated current $q$ is related to the tilted current $j$ via $q=\int_0^Tjdt$, combining $\eqref{eq;dmu}$ with this relation allows us to obtain the following relation between the SCGF and the variables $\rho$ and $H$:
\begin{align}\label{eq;CGFderivative}
\dv{\mu(\lambda)}{\lambda}=\int_0^Tdt\qty(-D(\rho)\partial_x\rho+\sigma(\rho)\partial_xH)\left.\right|_{x=0}.
\end{align}
From this relation, current SCGF is obtained.

\subsection{MFT equation for non-stationary current}
\vspace{5mm}
\begin{center}
\begin{tikzpicture}[
    node distance=1.5cm and 1.2cm,
    site/.style={circle, draw=black, thick, minimum size=10mm, inner sep=0pt, font=\small},
    reservoir/.style={rectangle, draw=black, thick, rounded corners, minimum width=2.2cm, minimum height=3cm, align=center, font=\small},
    arrow/.style={-{Stealth}, thick},
    hopping/.style={-{Stealth}, bend left=45, looseness=1.2, shorten >=1pt, shorten <=1pt},
    label_font/.style={font=\small},
    particle/.style={circle, fill=black!70, inner sep=0pt, minimum size=4pt},scale=0.8, transform shape
]

    \node[reservoir] (left_res) {Left\\Reservoir\\(Density $\rho_L$)};
    \node[site, right=1.2cm of left_res] (s1) {1};
    \node[site, right=of s1] (s2) {2};
    \node[right=0.5cm of s2] (dots1) {$\dots$};
    \node[site, right=0.5cm of dots1] (si) {$i$};
    \node[site, right=of si] (sj) {$j$};
    \node[right=0.5cm of sj] (dots2) {$\dots$};
    \node[site, right=0.5cm of dots2] (sN) {$N$};
    \node[reservoir, right=1.2cm of sN] (right_res) {Right\\Reservoir\\(Density $\rho_R$)};

    \node[particle] at ([yshift=-0.25cm]s1.center) {};
    \node[particle] at ([xshift=-0.2cm, yshift=-0.15cm]s2.center) {};
    \node[particle] at ([xshift=0.2cm, yshift=-0.15cm]s2.center) {};
    \node[particle] at ([yshift=0.25cm]s2.center) {};
    \node[particle] at ([xshift=-0.15cm, yshift=0.2cm]si.center) {};
    \node[particle] at ([xshift=0.15cm, yshift=-0.2cm]si.center) {};
    \node[particle] at ([xshift=-0.2cm]sN.center) {};
    \node[particle] at ([xshift=0.2cm]sN.center) {};

    \draw[arrow, line width=1.5pt] ([yshift=5mm]left_res.east) -- (s1.north west) node[midway, above] {$c_1^+$};
    \draw[arrow, line width=1.5pt] (s1.south west) -- ([yshift=-5mm]left_res.east) node[midway, below] {$c_1^-$};
    \draw[arrow, line width=1.5pt] ([yshift=5mm]right_res.west) -- (sN.north east) node[midway, above] {$c_N^+$};
    \draw[arrow, line width=1.5pt] (sN.south east) -- ([yshift=-5mm]right_res.west) node[midway, below] {$c_N^-$};

    \draw[hopping] (s1) to node[above, label_font] {$c^{1,2}$} (s2);
    \draw[hopping] (s2) to node[below, label_font] {$c^{2,1}$} (s1);
    \draw[hopping] (si) to node[above, label_font] {$c^{i,j}$} (sj);
    \draw[hopping] (sj) to node[below, label_font] {$c^{j,i}$} (si);
    \draw[hopping, shorten >=5pt] (s2) to (dots1);
    \draw[hopping, shorten <=5pt] (dots1) to (si);
    \draw[hopping, shorten >=5pt] (sj) to (dots2);
    \draw[hopping, shorten <=5pt] (dots2) to (sN);

\end{tikzpicture}
\end{center}
\vspace{4mm}

In this section, we derive the MFT equations for the integrated current at the left reservoir during the relaxation process toward the steady state. We consider stochastic lattice gases \cite{Bertini2002,Bertini2005,Bertini2006,Bertini2015}. This is a continuous-time Markov jump process of particles hopping on a one-dimensional lattice of $N$ sites, with the left and right ends connected to reservoirs with densities $\rho_L$ and $\rho_R$, respectively (macroscopic system length is $L$). The transition rate for a particle moving from site $i$ to site $j$ is denoted by $c^{ij}$. The particle exchange at the boundaries is modeled as the creation and annihilation of particles at sites 1 and $N$, with rate $c_1^{\pm}$ and $c_N^{\pm}$ respectively. Let $\eta\in\Omega$ denote the system configuration in the state space $\Omega$ ($\eta$ is given by the collection of site $i$ occupation variables $n_i$ ), and $f:\Omega\rightarrow\mathbb{R}$ be a observable. The generator $L$ of the Markov process is defined as follows:
\begin{align}\label{eq;Markovgenerator}
  Lf(\eta)=\sum_{|i-j|=1}c^{i,j}(f(\sigma^{ij}\eta)-f(\eta))+\sum_{i=1,N}c_i^{\pm}(f(\sigma_{\pm}^i\eta)-f(\eta))
\end{align}
Here, $\sigma^{ij}\eta$ represents the configuration after a particle hops from site $i$ to $j$, while $\sigma_{\pm}^i\eta$ denotes the configuration resulting from a particle creation or annihilation at the boundary site $i=1,N$.
Generally, the boundary rates $c_{1,N}^{\pm}$ are functions of the reservoir density $\rho_L$ and the system state $\eta$ (see Table $\ref{tab:rate}$ for the forms of specific models). For the transition rates, typical stochastic lattice gases such as the Symmetric Simple Exclusion Process (SEP), the Independent Random Walk (IRW), and the Zero Range Process (ZRP) take the following values \cite{Bertini2015}:

\begin{table}[H]
  \centering
  \caption{Example of transition rates}
  \label{tab:rate}
  \vspace{5pt}
  \begin{minipage}[t]{0.52\textwidth}
    \vspace{0pt} 
    \centering
    \begin{tabular}{c|ccc} 
          & $c^{i,j}$ & $c_1^+$ & $c_1^-$ \\ \hline
        SEP & $n_i(1-n_j)$ & $\Gamma\rho_L(1-n_1)$ & $\Gamma(1-\rho_L) n_1$\\
        IRW & $n_i$ & $\Gamma\rho_L$ & $\Gamma n_1$ \\
        ZRP & $g(n_i)$ & $\Gamma e^{\mu(\rho_L)}$ & $\Gamma g(n_1)$
    \end{tabular}
  \end{minipage}%
  \hfill 
  \begin{minipage}[t]{0.45\textwidth}
    \vspace{0pt} 
    \small 
    $n_i$ denotes the occupancy of site $i$. 
    For SEP and RBM, the parameters $\Gamma$ represents the boundary speed.
    The function $g(n_i)$ is the transition rate at which a particle moves from site $i$ to $i \pm 1$.
  \end{minipage}
\end{table}
To ensure that the system satisfies the local detailed balance condition\cite{Bertini2006,Bertini2007}, $c_1^-$ and $c_1^+$ are related to the chemical potential of the reservoir $\mu(\rho_L)$ and system Hamiltonian $\mathcal{H}$ via the relation 
\begin{align}\label{eq;LDB}
  \frac{c_1^+(\eta)}{c_1^-(\sigma_+^1\eta)}=\exp\qty{\mathcal{H}(\eta)-\mathcal{H}(\sigma_+^1\eta)+\mu(\rho_L)}.
\end{align}
$c_N^{\pm}$ also satisfy the same relation. In the bulk, $c^{i,i+1}$ and $c^{i+1,i}$ satisfy 
\begin{align}
  \frac{c^{i,i+1}(\eta)}{c^{i+1,i}(\sigma^{i,i+1}\eta)}=\exp\qty{\mathcal{H}(\eta)-\mathcal{H}(\sigma^{i,i+1}\eta)}.
\end{align}

To describe the macroscopic behavior of these stochastic lattice gases, we now take the hydrodynamic limit by applying the diffusive scaling.
In general, the boundary rates scale with the system size as $\Lambda^{-\theta}$ 
\cite{Baldasso2017,goncalves_non-equilibrium_2020,Saha2023}; they are expressed as $c^{\pm}_1 = C^{\pm}_1 \Lambda^{-\theta}$ and $c^{\pm}_N = C^{\pm}_N \Lambda^{-\theta}$, respectively. Here, $C_1^{\pm}$ and $C_N^{\pm}$ are macroscopic boundary rates that are independent of $\Lambda$. In the hydrodynamic limit $\Lambda \to \infty$, the system exhibits qualitatively different macroscopic behaviors corresponding to 
three distinct regimes: $\theta<1$, $\theta=1$, and $\theta>1$ \cite{Baldasso2017,Saha2023}.

The case $\theta<1$ is referred to as the fast-boundary regime. This corresponds to the case where the time scale of particle exchange at the boundaries is significantly shorter than the time scale of particle transport in the bulk. In this regime, boundary fluctuations become negligible, and the statistics of the integrated current $Q_T$ are independent of rate $C^{\pm}_1$ and $C_N^{\pm}$, only dependent on the reservoir density.

The case $\theta>1$ is referred to as the slow-boundary regime. This corresponds to a situation where the time scale of particle exchange at the boundaries is much longer than the time scale of particle transport within the bulk. Because the injection and removal of particles occur much more slowly than the internal relaxation of the system, the bulk can be considered to relax instantaneously and remain in a state of equilibrium. Consequently, the macroscopic current fluctuations are dominated by the processes at the boundaries.

 The case $\theta=1$ is referred to as the marginal-boundary regime, where boundary fluctuations can no longer be neglected. In this marginal regime, a density discontinuity generally exists at the interface between the reservoir and the system \cite{Derrida2021}. The scale dependence of the particle injection and removal rates is explicitly expressed as follows:
\begin{align}
  c^{\pm}_1 =\frac{C^{\pm}_1}{\Lambda}, \qquad c^{\pm}_N =\frac{C^{\pm}_N}{\Lambda}.
\end{align}
Here, the limits $C^{\pm}_1,\ C^{\pm}_N \to \infty$ and $C^{\pm}_1,\ C^{\pm}_N \to 0$ formally correspond to the fast-boundary and slow-boundary regimes, respectively. As exemplified in Table~$\ref{tab:rate}$, the fast-boundary and slow-boundary regimes are retrieved by taking the limits 
$\Gamma \to \infty$ and $\Gamma \to 0$ for the boundary speed parameter $\Gamma$, respectively. 
Since both regimes can be realized as asymptotic limits of the marginal-boundary case, we shall henceforth proceed with this marginal scaling.

In the hydrodynamic limit $\Lambda\rightarrow\infty$, the bulk transport coefficients $D(\rho)$ and $\sigma(\rho)$ are defined as follows: Assuming that the system is in local equilibrium, we denote the expectation value with respect to the local equilibrium distribution at density $\rho$ as $\mathbb{E}_{P^{\rho}_{eq}}[\cdot]$. To ensure the tractability of the hydrodynamic limit, we assume the gradient condition, where the current can be expressed as the gradient of a scalar local function $h(\eta)$. That is, $c^{i,i+1}-c^{i+1,i}$ can be written as $h(\tau_i\eta)-h(\tau_{i+1}\eta)$, where $\tau_i$ denotes the shift operator defined as $[\tau_i\eta](j)=\eta(j-i)$.
By defining $\Phi(\rho)\equiv\mathbb{E}_{P^{\rho}_{eq}}[h(\eta)]$,
the explicit forms of $D(\rho)$ and $\sigma(\rho)$ under the gradient condition are given as follows:
\begin{align}
  D(\rho)&\equiv\Phi'(\rho) \label{eq:D_gradient}\\
  \sigma(\rho)&\equiv\mathbb{E}_{P^{\rho}_{eq}}[c^{i,i+1}+c^{i+1,i}] \label{eq:sigma_gradient}
\end{align}
Under these definitions, the verification that the relations $\eqref{eq:Drho}$ and $\eqref{eq:sigmarho}$ are satisfied is provided in Appendix $\ref{app:A1}$.

Next, we describe the initial conditions. Two types of initial conditions are considered: the annealed initial condition, where initial particle positions are fluctuating (accounting for thermal fluctuations), and the quenched initial condition, where the initial configuration is fixed. In infinite systems, the difference in these initial fluctuations persists indefinitely and continues to influence the system dynamics\cite{Banerjee2022}. In finite systems, the system eventually settles into a unique non-equilibrium steady state regardless of the initial conditions, dependence on initial conditions is expected to vanish over time. Let $\rho(x,0)$ be the initial density distribution after taking the hydrodynamic limit. In the case of annealed initial conditions, we characterize its asymptotic behavior in the hydrodynamic limit as $P\qty(\rho(x,0))\asymp e^{-\Lambda\mathcal{F}[\rho(x,0)]}$. Here $\mathcal{F}$ denotes the large deviation functional for density fluctuations, which can be written in the form $\mathcal{F}[\rho(x,0)]=\int_0^L\mathcal{G}(\rho(x,0))dx$. In general, $\mathcal{G}(\rho(x,0))$ is the non-local functional of the density profile $\rho(x,0)$ \cite{Derrida2001,Derrida2007}. However, when the system is in equilibrium, it can be expressed as a local function in terms of the free energy functional $f(\rho)$ as follows\cite{Derrida2007}:
\begin{align}
  \mathcal{F}_{eq}(\rho(x,0))=\int_0^L dx\qty{f(\rho)-f(\brho)-(\rho-\brho)f'(\brho)}=\int_0^L dx\int_{\brho}^{\rho}dr\frac{2D(r)(\rho-r)}{\sigma(r)},
\end{align}
where $\brho$ is the equilibrium expectation value of the density.

By deriving the path integral representation of the moment-generating function for the integrated current at the origin on the lattice, and subsequently taking the hydrodynamic limit, we obtain the following MFT action\cite{Saha2023,Saha2024} (For the details of the calculation, see Appendix $\ref{ap;MFT action}$).

\begin{align}\label{eq;MFTaction}
  S[\rho,H]=\mathcal{F}[\rho(x,0)]-\int_0^Tdt\qty[\int_0^LH(x,t)\pdv{\rho(x,t)}{t}dx-\qty(\mathcal{H}_{bdry}^{(L)}+\mathcal{H}_{bulk}+\mathcal{H}_{bdry}^{(R)})], 
\end{align}
with MFT Hamiltonians that are defined as follows:
\begin{subequations}
\begin{align}
  \mathcal{H}_{bdry}^L&=C_1^+(\rho(0,t),\rho_L)\left(e^{\lambda+H(0,t)}-1\right) +C_1^-(\rho(0,t),\rho_L)\left(e^{-\lambda-H(0,t)}-1\right),\\
  \mathcal{H}_{bulk}&=\int_0^Ldx \ \qty(\frac{\sigma(\rho)}{2}\partial_xH-D(\rho)\partial_x\rho)\partial_xH,\\
  \mathcal{H}_{bdry}^R&=C_N^-(\rho(L,t),\rho_R)\left(e^{-H(L,t)}-1\right)+C_N^+(\rho(L,t),\rho_R)\left(e^{H(L,t)}-1\right).
\end{align}
\end{subequations}
Here, $C_1^{\pm}(\rho(0,t),\rho_L)$ and $C_N^{\pm}(\rho(L,t),\rho_R)$ represent the limiting forms of the boundary rates in the hydrodynamic limit. This corresponds to replacing the occupation number $n_1\ (n_N)$ at site 1 (site N) with the continuous density $\rho(0,t) (\rho(L,t))$ at the boundary. For the specific models introduced in Table~$\ref{tab:rate}$—namely, the SEP, IRW, and ZRP—these quantities are given as follows.
\begin{table}[H]
  \centering
  \caption{Example of boundary rates}
  \label{tab:ratemacro}
  \vspace{5pt}
    \begin{tabular}{c|ccc} 
          &  $C_1^{+}(\rho(0,t),\rho_L)$ & $C_1^{-}(\rho(0,t),\rho_L)$ \\ \hline
        SEP & $\Gamma\rho_L(1-\rho(0,t))$ & $\Gamma(1-\rho_L)\rho(0,t)$ \\
        IRW & $\Gamma\rho_L$ & $\Gamma \rho(0,t)$ \\
        ZRP & $\Gamma e^{\mu(\rho_L)}$ & $\Gamma g(\rho(0,t))$
    \end{tabular}
\end{table}

From the stationarity condition $\delta S[\rho,H]=0$, we obtain the following a set of MFT equations and boundary conditions. MFT equations are given as follows:
\begin{subequations}
\begin{align}
  \partial_t\rho&=\partial_x(D(\rho)\partial_x\rho-\sigma(\rho)\partial_xH),\label{eq;MFTeq1}\\
  \partial_tH&=-D(\rho)\partial_x^2H-\frac{\sigma'(\rho)}{2}(\partial_xH)^2. \label{eq;MFTeq2}
\end{align}
\end{subequations}
$\eqref{eq;MFTeq1}$ implies that the tilted current is expressed as $j=-D(\rho)\partial_x\rho+\sigma(\rho)\partial_xH$.
Boundary conditions at $x=0$ are
\begin{subequations}
\begin{align}
  &\sigma(\rho)\partial_xH-D(\rho)\partial_x\rho=C_1^+(\rho(0,t),\rho_L)e^{\lambda+H(0,t)}-C_1^-(\rho(0,t),\rho_L)e^{-\lambda-H(0,t)}\label{eq;leftmarginal1},\\
  &D(\rho)\partial_xH=-\fdv{C_1^+(\rho(0,t),\rho_L)}{\rho(0,t)}\left(e^{\lambda+H(0,t)}-1\right) -\fdv{C_1^-(\rho(0,t),\rho_L)}{\rho(0,t)}\left(e^{-\lambda-H(0,t)}-1\right)\label{eq;leftmarginal2},
\end{align}
\end{subequations}
and boundary conditions at $x=L$ are
\begin{subequations}
\begin{align}
  &\sigma(\rho)\partial_xH-D(\rho)\partial_x\rho=-C_N^+(\rho(L,t),\rho_R)e^{H(L,t)}+C_N^-(\rho(L,t),\rho_R)e^{-H(L,t)},\\
  &D(\rho)\partial_xH=\fdv{C_N^+(\rho(L,t),\rho_R)}{\rho(L,t)}\left(e^{H(L,t)}-1\right) +\fdv{C_N^-(\rho(L,t),\rho_R)}{\rho(L,t)}\left(e^{-H(L,t)}-1\right).
\end{align}
\end{subequations}
The initial conditions for $H$, for the case with thermal fluctuations (annealed initial condition) and the case with a fixed initial density (quenched initial condition), are given as follows respectively:
\begin{subequations}
\begin{align}
  &\text{Annealed initial condition}\quad H(x,0)=\int_0^Ldy\fdv{\mathcal{G}(y,0)}{\rho(x,0)},\qquad H(x,T)=0\\
  &\text{Quenched initial condition}\quad  H(x,T)=0
\end{align}
\end{subequations}

In the fast-boundary regime,$C_{1,N}^{\pm}\rightarrow\infty$ (more strictly, $\Gamma\rightarrow\infty$), the interface with the reservoir relaxes to equilibrium instantaneously, thereby fixing the boundary density to that of the reservoir, $\rho_L=\rho(0,t)$ \cite{Derrida2021}.\footnote{In Section 3, we only treat fast-boundary. In Section 4, we also treat outside the fast-boundary regime.}
Using the boundary conditions $\eqref{eq;leftmarginal1}$, the following can be demonstrated. By factoring out a parameter $\Gamma$, that represents the boundary speed, such that $C_1^{\pm}=\Gamma \widetilde{C}_1^{\pm}$, the condition for the hydrodynamic current $j=\Gamma(\widetilde{C}_1^+(\rho(0,t),\rho_L)-\widetilde{C}_1^-(\rho(0,t),\rho_L))$ to remain finite in the limit $\Gamma\rightarrow\infty$ requires that $\widetilde{C}_1^+(\rho(0,t),\rho_L)=\widetilde{C}_1^-(\rho(0,t),\rho_L)$. From this requirement and the condition that the right-hand side of $\eqref{eq;leftmarginal1}$ does not diverge as $\Gamma\rightarrow\infty$, we obtain $H(0,t)=-\lambda$.
Meanwhile, the local detailed balance condition $\eqref{eq;LDB}$ in the hydrodynamic limit is expressed as follows:
\begin{align}
  \frac{\widetilde{C}_1^+(\rho(0,t),\rho_L)}{\widetilde{C}_1^-(\rho(0,t),\rho_L)}=\exp[\mu(\rho_L)-\mu(\rho(0,t))]
\end{align}
Therefore, in the fast-boundary limit $(\Gamma\rightarrow\infty)$, we obtain $\mu(\rho(0,t))=\mu(\rho_L)$, which implies $\rho(0,t)=\rho_L$.
By applying a similar argument to the boundary on the opposite side, in the fast-boundary limit $C_{1,N}^{\pm}\rightarrow\infty$, the complex boundary conditions reduce to the following Dirichlet forms:
\begin{align*}
  \rho(0,t)=\rho_L,\  H(0,t)=-\lambda,\ \rho(L,t)=\rho_R,\ H(L,t)=0.
\end{align*}
For the SEP and RBM introduced above, it is directly evident from the boundary equations $\eqref{eq;leftmarginal1}$ and $\eqref{eq;leftmarginal2}$ that they reduce to the Dirichlet boundary conditions in the fast-boundary limit $\Gamma\rightarrow\infty$, without requiring the discussion above. In the case of the SEP, the equations $\eqref{eq;leftmarginal1}$ and $\eqref{eq;leftmarginal2}$ take the following form:
\begin{subequations}
\begin{align}
  &\sigma(\rho)\partial_xH-D(\rho)\partial_x\rho=\Gamma \qty{\rho_L(1-\rho(0,t)) e^{\lambda+H(0,t)}-\rho(0,t)(1-\rho_L) e^{-\lambda-H(0,t)}}\\
  &D(\rho)\partial_xH=-\Gamma\qty{\rho_L\left(e^{\lambda+H(0,t)}-1\right) +(1-\rho_L)\left(e^{-\lambda-H(0,t)}-1\right)}
\end{align}
\end{subequations}
The condition for the right-hand side to be finite as $\Gamma\rightarrow\infty$ imposes the boundary conditions $\rho(0,t)=\rho_L$ and $H(0,t)=-\lambda$\cite{Saha2023}. A similar argument holds for the boundary at $x=L$.

\section{Current variance for general $\sigma(\rho)$}\label{general}
\subsection{Model and formalism}
In this section, we derive the variance of the integrated current at the left reservoir boundary for a constant diffusion coefficient $D$ and arbitrary mobility $\sigma(\rho)$ under the fast-boundary condition. Regardless of the initial and boundary conditions,
the MFT equations are given as follows: 
\begin{subequations}\label{eq;MFTgeneral}
\begin{align}
   \partial_t{\rho}&=D\partial_x^2{\rho}-\partial_{x}\qty(\sigma(\rho)\partial_x{H}),\\
  \partial_t{H}&=-D\partial_x^2{H}-\frac{\sigma'(\rho)}{2}\qty(\partial_x{H})^2.
\end{align}
\end{subequations}
We assume that the bulk is initially in a uniform state with mean density $\brho$. However, depending on the presence of initial fluctuations, two distinct types of initial conditions can be considered: the annealed initial condition, which accounts for thermal fluctuations around $\brho$, and the quenched initial condition, which the initial density profile is fixed to its average $\brho$ without any fluctuations. The large deviation function $\mathcal{F}(\rho)$ is
\begin{align}
  \mathcal{F}(\rho)=\begin{cases}
    \int_0^Ldx\int_{\bar{\rho}}^{\rho(x,0)}dr\frac{2D\qty(\rho(x,0)-r)}{\sigma(r)}\qquad &\text{Annealed initial condition}\\
    0\qquad &\text{Quenched initial condition}.
  \end{cases}
\end{align}
Since the MFT equations $\eqref{eq;MFTgeneral}$ are generally a set of coupled non-linear partial differential equations for $\rho$ and $H$, it is not possible to obtain a general solution. However, for a constant diffusion coefficient $D$ independent of $\rho$, cumulants of arbitrary order can be calculated through perturbative expansion\cite{Krapivsky2012}. In the following, we calculate the current variance by performing a perturbation expansion with respect to the counting field $\lambda$. From the definition of cumulants,
\begin{align}\label{eq;cgfeapand}
  \mu(\lambda)=\lambda\ev{Q_T}_c+\frac{\lambda^2}{2!}\ev{Q_T^2}_c+\frac{\lambda^3}{3!}\ev{Q_T^3}_c+\cdots,
\end{align}
and substituting $\eqref{eq;cgfeapand}$ into $\eqref{eq;CGFderivative}$, $\rho$ and $H$ satisfy following relation
\begin{align}\label{eq;3-14}
  \ev{Q_T}_c+\lambda \ev{Q_T^2}_c+\cdots=\int_0^Tdt\qty(-D\partial_x\rho+\sigma(\rho)\partial_xH)\left.\right|_{x=0}
\end{align}
We expand $\rho$ and $H$ in powers of $\lambda$ 
\begin{subequations}
\begin{align}\label{eq:p-lambda}
  \rho&=\rho_0+\lambda \rho_1+\lambda^2 \rho_2+\cdots,\\
  H&=\lambda H_1+\lambda^2 H_2+\cdots,
\end{align}
\end{subequations}
and substitute these into $\eqref{eq;3-14}$. By comparing the terms of order $\lambda^0$ and $\lambda$, we obtain the following expressions for the first and second cumulants:
\begin{align}
  \ev{Q_T}_c&=-\int_0^TdtD\partial_x\rho_0\left.\right|_{x=0},\\
  \ev{Q_T^2}_c&=\int_0^Tdt\qty{-D\partial_x\rho_1+\sigma(\rho_0)\partial_xH_1}\left.\right|_{x=0} \label{eq;variance}.
\end{align}
Thus, to obtain the current variance $\ev{Q^2}_c$, we need to determine $\rho_0$, $\rho_1$ and $H_1$.

\subsection{Result}
\subsubsection*{Annealed initial condition}
In the case of annealed initial condition, temporal boundary conditions are 
\begin{align}\label{eq;tbdry}
  H(x,0)=f'(\rho)-f'(\bar{\rho}) ,\qquad H(x,T)=0,
\end{align}
and spatial boundary conditions are 
\begin{align}\label{eq;sbdry}
  H(0,t)=-\lambda, \qquad \rho(0,t)=\rho_L \qquad  H(L,t)=0, \qquad \rho(L,t)=\rho_R.
\end{align}
$f(\rho)$ is the free energy density. Performing a perturbative expansion of the MFT equations $\eqref{eq;MFTgeneral}$ subject to the specified boundary conditions $\eqref{eq;tbdry}$ $\eqref{eq;sbdry}$, we obtain the analytical expressions for $\rho_0(x,t)$ and $H_1(x,t)$ (Details are given in Appendix $\ref{ap_gene}$):
\begin{align}\label{eq;rho}
  \rho_0(x,t)&=\frac{\rho_R-\rho_L}{L}x+\rho_L+\sum\limits_{n=1}^{\infty}\frac{2}{n\pi}\qty{-\rho_L+\rho_R(-1)^n+\bar{\rho}(1-(-1)^n)}e^{-D\qty(\frac{n\pi}{L})^2t}\sin\frac{n\pi}{L}x,\\
  H_1(x,t)&=\frac{x}{L}-1+\sum\limits_{n=1}^{\infty}\frac{2}{n\pi}e^{-D\qty(\frac{n\pi}{L})^2(T-t)}\sin\frac{n\pi}{L}x.\label{eq;H}
\end{align}
To express the current variance in terms of $\rho_0$ and $H_1$, we introduce the Green's function $G^D(x,y;t)$ for the diffusion equation on the interval $0<x<L$ with Dirichlet boundary conditions:
\begin{align}\label{eq;GreenDirichlet}
  G^D(x,y;t)=\frac{2}{L}\sum\limits_{n=1}^{\infty}\sin\qty(\frac{n\pi}{L}x)\sin\qty(\frac{n\pi}{L}y)e^{-D\qty(\frac{n\pi}{L})^2t}.
\end{align}
Using this Green's function, current variance is obtained as follows :
\begin{align}\label{eq;generalanneal}
\begin{split}
  \ev{Q_T^2}_{c,\mathcal{A}}&=-\frac{1}{2}\int_0^Tdt\int_0^LdyG^D_x(x=0,y;t)\sigma(\rho_0(y,0))H_1(y,0)\\
  &\qquad-\int_0^Tdt\int_0^Ldy\int_0^{t}dsDG^D_{xy}(x=0,y;t-s)(\sigma(\rho_0(y,s))\partial_y H_1(y,s))
  +\int_0^Tdt\sigma(\rho_0)\partial_xH_1\left.\right|_{x=0}.
\end{split}
\end{align}
Here, the first term represents the effect of initial fluctuations. The third term captures the direct local fluctuation at the boundary $x=0$, where the product  $\sigma(\rho)\partial_xH_1$ denotes the current fluctuation driven by the effective external field $\partial_xH_1(x=0,t)$. The second term describes the propagation of fluctuations from position $y$ and time $s$ in the bulk to the left boundary at time $t$. 
By rearranging these terms, we obtain the following compact form (Details are given in Appendix $\ref{ap_gene}$):
\begin{align}\label{eq;generalanneal2}
  \ev{Q_T^2}_{c,\mA}=\int_0^Tdt\int_0^Ldy \sigma(\rho_0)(\partial_yH_1)^2+\int_0^Ldy\frac{\sigma(\bar{\rho})}{2D}(H_1(y,0))^2.
\end{align}

\subsubsection*{Quenched initial condition}
In the case of quenched initial condition, temporal boundary condition is 
\begin{align}
   H(x,T)=0,
\end{align}
and spatial boundary conditions are 
\begin{align}
  H(0,t)=-\lambda, \qquad \rho(0,t)=\rho_L,\qquad
  H(L,t)=0, \qquad \rho(L,t)=\rho_R.
\end{align}
There is no constraint on $H(x,0)$. From these boundary conditions and perturbative calculation (Details are given in Appendix $\ref{ap_gene}$), current variance is given by
\begin{align}\label{eq;generalquench}
  \ev{Q_T^2}_{c,\mathcal{Q}}=-\int_0^Tdt\int_0^Ldy\int_0^{t}dsDG^D_{xy}(x=0,y;t-s)(\sigma(\rho_0(y,s))\partial_y H_1(y,s))+\int_0^Tdt\sigma(\rho_0)\partial_xH_1\left.\right|_{x=0}.
\end{align}
$\rho_0$, $H_1$ and $G^D(x,y;t)$ are same as annealed case $\eqref{eq;rho}$, $\eqref{eq;H}$ and $\eqref{eq;GreenDirichlet}$ respectively. Similarly to the annealed case, the first term represents the propagation of fluctuations from the bulk, while the second term captures the direct local fluctuation at the boundary $x=0$. Rearranging these terms, we obtain the following compact form:
\begin{align}\label{eq;generalquench2}
 \ev{Q_T^2}_{c,\mQ}=\int_0^Tdt\int_0^Ldy\, \sigma(\rho_0)(\partial_yH_1)^2.
\end{align}

In this way, by combining the given model parameters (reservoir densities, initial condition, system length, diffusion coefficient, and mobility) with $\eqref{eq;rho}$ and $\eqref{eq;H}$, the current variance for both annealed and quenched initial conditions can be evaluated from $\eqref{eq;generalanneal2}$ and $\eqref{eq;generalquench2}$ respectively. For example, by setting $D=1$ and $\sigma(\rho)=2\rho(1-\rho)$ in the above result, we obtain the current fluctuations for the SEP; similarly, setting $D=1$ and $\sigma(\rho)=\rho^2$ yields those for the KMP model\cite{Kipnis1982}.

From the above results $\eqref{eq;generalanneal2}$ and $\eqref{eq;generalquench2}$, $\ev{Q_T^2}_{c,\mA}$ and $\ev{Q_T^2}_{c,\mQ}$ satisfy the following relation:
\begin{align}
   \ev{Q_T^2}_{c,\mA}=\ev{Q_T^2}_{c,\mQ}+\int_0^Ldy\frac{\sigma(\bar{\rho})}{2D}(H_1(y,0))^2.
\end{align}
This explicitly shows that the effect of initial fluctuations is decoupled and simply adds to the quenched variance. Since the second term on the right-hand side is obviously positive, the inequality $\ev{Q_T^2}_{c,\mA}>\ev{Q_T^2}_{c,\mQ}$ holds. We expect the current variance to be larger for the annealed initial condition than for the quenched one, as the former includes initial fluctuations.

Moreover, these expressions are expected to correspond to the macroscopic counterpart of the dynamical fluctuation-response relations derived by \cite{aslyamov_dynamical_2026}, where the integrated observables are regarded as the integrated current.

\subsubsection*{Example}
Here, we show the example of current variance for SEP($D(\rho)=1,\ \sigma(\rho)=2\rho(1-\rho)$) and IRW ($D(\rho)=1,\ \sigma(\rho)=2\rho$). The following figure shows the current variance calculated from $\eqref{eq;generalanneal2}$ and $\eqref{eq;generalquench2}$.
\begin{figure}[H]
  \centering 
  \begin{minipage}[b]{0.48\textwidth}
    \quad\textbf{(a)} 
  \end{minipage}
  \hfill
  \begin{minipage}[b]{0.48\textwidth}
    \quad\textbf{(b)} 
  \end{minipage}
  \vspace{3pt}
  \begin{minipage}[b]{0.48\textwidth}
    \centering
    \includegraphics[width=\textwidth]{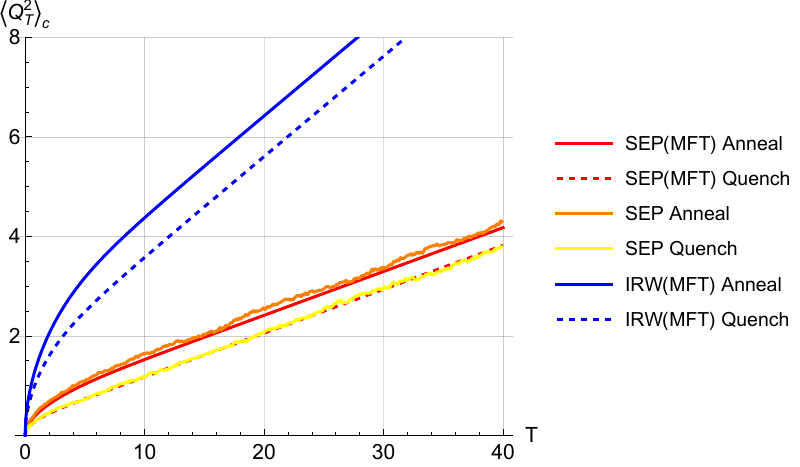}
  \end{minipage}
  \hfill
  \begin{minipage}[b]{0.48\textwidth}
    \centering
    \includegraphics[width=\textwidth]{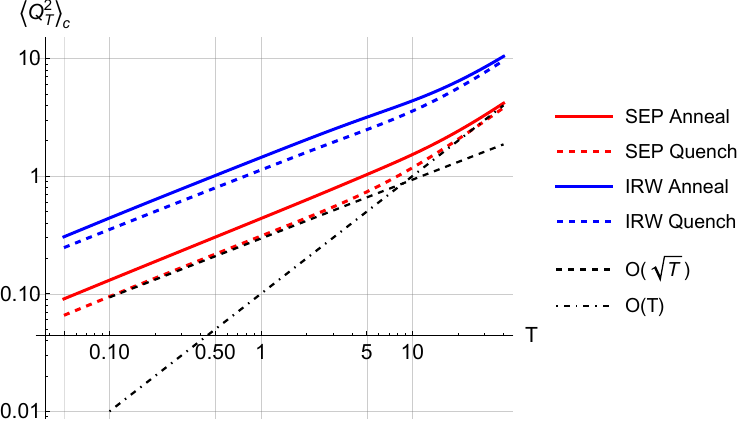}
  \end{minipage}

  \vspace{5pt}

\caption{\textbf{(a)}Calculated results for the integrated current variance $\ev{Q_T^2}_c$ using $\eqref{eq;generalanneal2}$ $\eqref{eq;generalquench2}$ with parameters $L=5,\ \rho_L=0.8, \ \rho_R=0.2, \ \brho=0.5$. For the SEP, Monte Carlo simulation results---performed with scaling parameter $\Lambda=150$ (system size $N=750$), boundary coupling strength $\Gamma=10^4$, and $4000$ ensemble samples---are overlaid for comparison. \textbf{(b)}Log-log plot of the same data shown in (a) with the asymptotic line.}
\label{fig:variance_comparison}
\end{figure}
Both models exhibit a crossover from $O(\sqrt{T})$ to $O(T)$. In the regime $T\ll L^2$, interference between the two reservoirs is negligible, and the integrated current is expected to behave similarly to that of a semi-infinite system with a single reservoir. Conversely, for $T\gg L^2$, the system should approach a non-equilibrium steady state. Consequently, the $n$-th cumulant of the integrated current is anticipated to exhibit a crossover from $O(\sqrt{T})$ to $O(T)$.  
The fundamental difference between SEP and IRW lies in the presence or absence of the volume exclusion effect. While both model exhibit a crossover from $O(\sqrt{T})$ to $O(T)$, the growth rate in the SEP is suppressed by the exclusion effect. The mobility is given by $\sigma(\rho)=2\rho(1-\rho)$ for SEP and $\sigma(\rho)=2\rho$ for IRW. Since $\sigma_{SEP}<\sigma_{IRW}$, this can be attributed to the fact that the larger response coefficient directly leads to enhanced fluctuations.

We can also check the validity of the MFT by comparing its analytical predictions with Monte Carlo simulations of the lattice model, specifically the SEP. As shown in Fig $\ref{fig:variance_comparison}(a)$, the quantitative agreement between the two is excellent, confirming that the MFT accurately captures the macroscopic fluctuations originating from the microscopic dynamics of the system.

\subsection{Consistency with previous result}
\subsection*{$L\rightarrow\infty$}
Our results are expected to recover the semi-infinite case, which connect a reservoir at $x=0$, in the large-system limit $L\rightarrow\infty$ while keeping the left reservoir fixed. For diffusive systems with a constant diffusion coefficient and arbitrary mobility, the current variance in an infinite system with a step initial condition has been derived using MFT in
\cite{Krapivsky2012}\footnote{The method in\cite{Krapivsky2012} is specific to infinite and semi-infinite geometries, which is distinct from our approach.}. By applying a similar approach, the current variance for a semi-infinite system is obtained as follows (The subscript $si$ denotes the results for the semi-infinite system. Details of derivation are given in Appendix $\ref{Appendix_si_cgf}$) :
\begin{align}
  \ev{Q_T^2}_{c,\mQ} &= \int_0^Tdt\int_0^{\infty}dx\sigma(\rho_0^{si})(\partial_xH_1^{si})^2 \label{eq:var_si_q} \\
  \ev{Q_T^2}_{c,\mA} &= \int_0^Tdt\int_0^{\infty}dy\sigma(\rho_0^{si})(\partial_xH_1^{si})^2+\frac{(2-\sqrt{2})\sigma(\brho)\sqrt{T}}{\sqrt{D\pi}} \label{eq:var_si_a}
\end{align}
One can verify that, in the limit $L\rightarrow\infty$, the results for both the annealed initial condition $\eqref{eq;generalanneal2}$ and quenched initial condition $\eqref{eq;generalquench2}$ reduce to $\eqref{eq:var_si_a}$ and $\eqref{eq:var_si_q}$ respectively (Details are given in Appendix $\ref{Appendix_fi_cgf}$).

\subsection*{Long time average}
The system relaxes to a steady state in the long-time limit. Therefore, average current of steady state is obtained as $\lim\limits_{T\rightarrow\infty}\frac{\ev{Q_T^2}_c}{T}$. Taking the long-time average of $\eqref{eq;generalanneal2}$ and $\eqref{eq;generalquench2}$, both the annealed and quenched cases reduce to the following  (Details of the derivation are given in Appendix $\ref{Appendix_ness}$):
\begin{align} \label{eq:long_time_av}
  \lim\limits_{T\rightarrow\infty}\frac{\ev{Q_T^2}_c}{T}=\frac{1}{L^2}\int_0^L\sigma(\rho^{st})dx.
\end{align}
Here, $\rho^{st}$ denotes the steady-state density profile.
This result is consistent with the one derived in \cite{Bodineau2004} using additivity principle. The expression for the current variance derived above is applicable to an arbitrary $\sigma(\rho)$ with a constant $D$, whereas Ref \cite{Bodineau2004} is limited to cases satisfying certain condition \cite{Bertini2006,Shpielberg2016}. It is suggested that as long as we focus on the second cumulant, it is possible to assume additivity principle.

\section{Current SCGF of RBM }\label{sectionRBM}

\subsection{Reflective Brownian Motion}\label{sec;model}

In this section, we consider Reflective Brownian Motions (RBM) on a one-dimensional finite system. The system consists of point particles undergoing Brownian motion subject to hard-core interactions, meaning that particles cannot pass each other. This non-passing property renders the system a prototype of single-file diffusion model \cite{Krapivsky2014,hegde_universal_2014,Krapivsky2015}.

\begin{figure}[H]
  \centering
  \begin{minipage}[c]{0.55\textwidth}
    \centering
    \includegraphics[width=\textwidth]{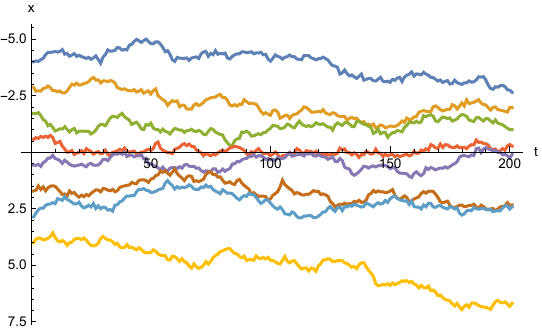}
  \end{minipage}
  \hfill 
  \begin{minipage}[c]{0.4\textwidth}
    \caption{An example of RBM trajectories }
    \label{fig:large deviation function(anneal)}
    \vspace{10pt}
  \end{minipage}
\end{figure}
The boundary behavior is defined analogously to the stochastic lattice gases introduced in Section 2. The system is connected to two particle reservoirs: a reservoir L with density $\rho_L$ at the left boundary $(x=0)$ and a reservoir R with density $\rho_R$ at the right boundary ($x=L$). At the left reservoir, particles are injected at rate $a$ and extracted at rate $c$. Similarly, at the right reservoir, particles are extracted at rate $b$ and injected at rate $d$. We assume the local detailed balance conditions $a/c=\rho_L$ and $d/b=\rho_R$. The single particle distribution function $P(x,t)$ evolves according to the diffusion equation $\partial_tP(x,t)=\partial_x^2P(x,t)$. The boundary conditions for $P(x,t)$ are given as follows:
\begin{subequations}
\begin{align}
  &x=0 \ \cdots \ \partial_xP(x=0,t)=c P(0,t)\\
  &x=L \ \cdots \ \partial_xP(x=L,t)=-b P(L,t)
\end{align}
\end{subequations}
Unlike the stochastic lattice gas models discussed in the previous section,  RBM is a stochastic model defined on a continuous space. To this end, we introduce a similar scaling transformation for such continuous space models; specifically, we define the macroscopic coordinates $(x,t)$ through the diffusive scaling $x=i/\Lambda$ and $t=\tau/\Lambda^2$ relative to the microscopic coordinates $(i,\tau)$ $(0<i<\Lambda L,\ 0<\tau<\Lambda^2 T)$ \cite{grabsch_exact_2025,grabsch_macroscopic_2026}. In the limit $\Lambda\rightarrow\infty$, the microscopic dynamics converges to the macroscopic description.

Since identical Brownian particles undergo elastic collisions in one dimension, their spatial configuration is equivalent to that of a system of non-interacting Brownian particles when particle labels are ignored. When focusing on observables that do not require particle labeling—such as the integrated current—the system of RBM can be treated as a collection of independent Brownian particles \cite{Harris1965}.\\
As before, we assume that the bulk is initially in a uniform state with mean density $\brho$. We consider two types of initial conditions: the annealed initial condition, which accounts for thermal fluctuations, and the quenched initial condition, where the Brownian particles position are fixed without any initial fluctuations. The large deviation function of $\rho(x,0)$ is
\begin{align}
  \mathcal{F}(\rho(x,0))=\begin{cases}
    \int_0^Ldx\int_{\bar{\rho}}^{\rho(x,0)}dr\frac{\rho(x,0)-r}{r}\qquad &\text{Annealed initial condition}\\
    0\qquad &\text{Quenched initial condition}
  \end{cases}
\end{align}

In terms of the density field $\rho(x,t)$, RBM can be characterized as a system with a diffusion coefficient $D(\rho)=1$, and mobility $\sigma(\rho)=2\rho$. This allows RBM to be interpreted as the low-density limit of the SEP---which has $D(\rho)=1$, $\sigma(\rho)=2\rho(1-\rho)$---by neglecting terms of $O(\rho^2)$.

\subsubsection*{Boundary conditions and scale transform}
As classified in Section 2.2, the boundary driving exhibits distinct physical regimes depending on the scaling exponent $\theta$. In this section, we calculate the current SCGF for the RBM 
under the marginal boundary regime, where the particle injection and removal rates are scaled as:
\begin{equation}
  a = \frac{A}{\Lambda}, \quad b = \frac{B}{\Lambda}, \quad
  c = \frac{C}{\Lambda}, \quad d = \frac{D}{\Lambda}.
\end{equation}
In the slow boundary limit ($A, B, C, D \to 0$), the current SCGF under the standard $O(\Lambda)$ scaling vanishes. This does not imply that the microscopic fluctuations strictly disappear, 
but rather that they are governed by a different large deviation speed that is slower than $\Lambda$. Consequently, the MFT analysis at the scale of $\Lambda$ cannot capture these sub-leading boundary fluctuations. In contrast, the marginal regime with finite parameters retains a non-trivial coupling between bulk diffusion and boundary fluctuations. As demonstrated in Section \ref{sec;fast}, our exact solution for this marginal regime also systematically reproduces the fast boundary results in the limit $A, B, C, D \to \infty$.

\subsubsection*{Remark}
In the marginal-boundary regime, mean integrated current $\ev{Q_T}$ exhibits the following behavior;

 \begin{figure}[H]
  \centering
  \begin{minipage}[c]{0.55\textwidth}
    \centering
    \includegraphics[width=\textwidth]{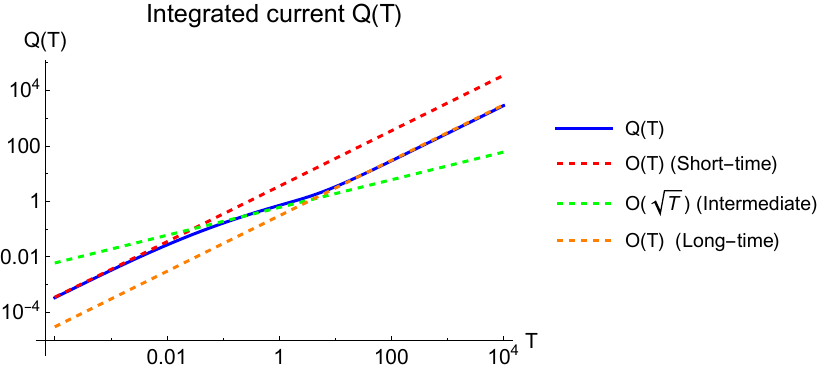}
  \end{minipage}
  \hfill 
  \begin{minipage}[c]{0.4\textwidth}
    \caption{Integrated current  with parameters $L=5,\ \rho_L=0.8, \ \rho_R=0.2, \ \brho=0.5$, $A=B=10$}
    \label{fig:large deviation function(anneal)}
    \vspace{10pt}
  \end{minipage}
\end{figure}
The integrated current initially scales as $O(T)$ before crossing over to $O(\sqrt{T})$ at $T\sim\frac{\rho_L^2}{A^2}$. It eventually returns to $O(T)$ scaling at $T\sim L^2$ at the diffusive time scale.
The marginal-boundary case leads to the emergence of an $O(T)$ scaling for $Q_T$ at early times $(T\lesssim \frac{\rho_L^2}{A^2})$. When rate $A$ is sufficiently large, the initial $O(T)$ regime is negligible; thus, this behavior is not a universal feature common to all finite boundary-driven diffusive systems.\\

Although RBM is defined on a continuous space, current large deviations are equivalent to the Independent Random Walk (IRW), one of the stochastic lattice gas model.
IRW is non-interacting particles undergoing simple random walks on a one-dimensional lattice, with both ends coupled to particle reservoirs at densities $\rho_L$ and $\rho_R$. In the bulk, each particle hops to the adjacent right or left site at a rate of 1. At the left boundary (site 1), particles are injected from the reservoir at rate $a$ and removed at rate $c$. Similarly, at the right boundary (site $N$), particles are removed at rate $b$ and injected at rate $d$. We assume the local detailed balance conditions $a/c=\rho_L$ and $d/b=\rho_R$
(The relationship between RBM and IRW is explained in Appendix $\ref{equivalence}$). Therefore, MFT formalism introduced in Section $\ref{section2}$ could be applied to the RBM by considering equivalent IRW .

\subsection{MFT analysis}\label{sec; MFT action}
As detailed in Appendix E, the MFT framework for the RBM reduces to that for IRW since they share the same bulk transport coefficients ($D(\rho)=1, \sigma(\rho)=2\rho$) and boundary rates. Substituting these specific coefficients into the general MFT equations $\eqref{eq;MFTeq1},\eqref{eq;MFTeq2}$, the governing equations for the RBM are explicitly given by
\begin{subequations}\label{eq;MFTRBM}
\begin{align}
     \partial_t\rho=\partial_x\qty(\partial_x\rho-2\rho\partial_xH)\\
     \partial_tH=-\partial_x^2H-\qty(\partial_xH)^2
\end{align}
\end{subequations}
By analytically solving these MFT equations subject to the prescribed boundary conditions, we determine the SCGF of the integrated current for both annealed and quenched initial condition.

\subsubsection*{Annealed initial condition}
From the general boundary conditions $\eqref{eq;leftmarginal1}$ and $\eqref{eq;leftmarginal2}$, boundary conditions for the RBM at $x=0$ are 
\begin{subequations}\label{eq;RBMbdryL}
\begin{align}
  \partial_xH(x=0,t)&=-\frac{A}{\rho_L}\qty(e^{-\lambda-H(0,t)}-1),\\
  2\rho\partial_xH(x=0,t)-\partial_x\rho(x=0,t)&=\frac{A}{\rho_L}\qty(\rho_Le^{\lambda+H(0,t)}-\rho(0,t)e^{-\lambda-H(0,t)}),
\end{align}
\end{subequations}
and boundary conditions at $x=L$ are
\begin{subequations}\label{eq;RBMbdryR}
\begin{align}
  \partial_xH(x=L,t)&=B\qty(e^{-H(L,t)}-1),\\
  2\rho\partial_xH(x=L,t)-\partial_x\rho(x=L,t)&=B\qty(\rho(L,t)e^{-H(L,t)}-\rho_Re^{H(L,t)}).
\end{align}
\end{subequations}
Temporal conditions are given by
\begin{align*}
  H(x,0)=\log\rho-\log\bar{\rho},\qquad H(x,T)=0.
\end{align*}
Although the MFT equations $\eqref{eq;MFTRBM}$ and their associated boundary conditions $\eqref{eq;RBMbdryL}$, $\eqref{eq;RBMbdryR}$ are inherently nonlinear and coupled, the specific transport coefficients of RBM ($D=1$ and $\sigma(\rho)=2\rho$) allow for an exact linearization. By introducing the Cole-Hopf transformation \cite{Derrida2009a},
\begin{align}
  P=e^H,\qquad Q=\rho e^{-H},
\end{align}
we can remarkably simplify the problem into a set of independent PDEs. Specifically, the $P(x,t)$ and $Q(x,t)$ satisfy backward and forward diffusion equations, 
\begin{align}\begin{cases}
   \partial_tQ=\partial_x^2Q\\
   \partial_tP=-\partial_x^2P.
\end{cases}
\end{align}
The spatial boundary conditions of $P(x,t)$ and $Q(x,t)$ are
\begin{equation}\label{eq;spaceboundary}
\begin{split}
  P(0,t)-\frac{\rho_L}{A}\pdv{P(x=0,t)}{x}&=e^{-\lambda},\quad P(L,t)+\frac{1}{B}\pdv{P(x=L,t)}{x}=1\\
  Q(0,t)-\frac{\rho_L}{A}\pdv{Q(x=0,t)}{x}&=\rho_Le^{\lambda},\quad Q(L,t)+\frac{1}{B}\pdv{Q(x=L,t)}{x}=\rho_R,
  \end{split}
\end{equation}
and temporal boundary conditions are given by
\begin{align}
  Q(x,0)=\bar{\rho},\qquad P(x,T)=1.
\end{align}
To facilitate the calculation, we introduce the Green's function $G(x,y;t)$ for the diffusion equation $\partial_tG=\partial_x^2G$, subject to the following Robin boundary conditions: $G(0,y;t)-\frac{\rho_L}{A}\pdv{G(0,y;t)}{x}=0$ and $G(L,y;t)+\frac{1}{B}\pdv{G(L,y;t)}{x}=0$.
The Green's function can be expanded in terms of the eigenfunctions $X_n(x)$ as follows:
\begin{align}\label{finite green fun}
  G^R(x,y;t)=\sum\limits_{n=1}^{\infty}\frac{X_n(x)X_n(y)}{N_n}e^{-k_n^2t}
\end{align}
where $X_n(x)$ represent the eigenfunction of the associated boundary value problem, and $N_n$ denote the normalization constant. These are explicitly given by
\begin{align}
  X_n(x)&\equiv\cos k_nx+\frac{A}{k_n\rho_L}\sin k_nx,\label{eq;Xn}\\
  N_n&\equiv\int_0^L|X_n(x)|^2dx,\label{eq;Nn}
\end{align}
where the eigenvalue $k_n$ are determined by the transcendental equation $(B+A/\rho_L)k_n=(k_n^2-AB/\rho_L)\tan(k_nL)$.\\
Using this Green's function $\eqref{finite green fun}$, $P(x,t)$ and $Q(x,t)$ are obtained as follows:
\begin{align}
  P(x,t)&=a_px+b_p+\int_0^LG^R(x,y;T-t)(1-a_py-b_p)dy, \label{eq;P}\\
  Q(x,t)&=a_qx+b_q+\int_0^LG^R(x,y;t)(\bar{\rho}-a_qy-b_q)dy.\label{eq;Q}
\end{align}
$a_p,\ a_q, \ b_p,\ b_q $ are the parameters that are defined as:
\begin{equation}\label{eq;a_p}
\begin{split}
  a_p&=\frac{1-e^{-\lambda}}{L+\frac{1}{B}+\frac{\rho_L}{A}}, \quad b_p=\frac{1-e^{-\lambda}}{1+\frac{AL}{\rho_L}+\frac{A}{B\rho_L}}+e^{-\lambda},\\
  a_q&=\frac{\rho_R-\rho_Le^{\lambda}}{L+\frac{1}{B}+\frac{\rho_L}{A}}, \quad b_q=\frac{\rho_R-\rho_Le^{\lambda}}{1+\frac{AL}{\rho_L}+\frac{A}{B\rho_L}}+\rho_Le^{\lambda}.
\end{split}
\end{equation}
From $\eqref{eq;CGFderivative}$, $\eqref{eq;P}$, and $\eqref{eq;Q}$, the current SCGF is obtained as follows: 
\begin{align}\label{cumulant MFT}
\begin{split}
  \mu_{\mA}(\lambda) &=\int_0^{\lambda}d\lambda'\int_0^Tdt(Q\partial_xP-P\partial_xQ)\left.\right|_{x=0}\\
  &= \left[ \frac{\rho_L T}{L + \frac{1}{B} + \frac{\rho_L}{A}} + \sum_{n=1}^{\infty} \frac{1 - e^{-k_n^2 T}}{N_n k_n^2} \left\{ -\frac{A}{L + \frac{1}{B} + \frac{\rho_L}{A}} S_1 + \left( A - \frac{\rho_L}{L + \frac{1}{B} + \frac{\rho_L}{A}} \right) S_2 \right\} \right] (e^{\lambda} - 1) \\
  &\quad + \left[ \frac{\rho_R T}{L + \frac{1}{B} + \frac{\rho_L}{A}} + \sum_{n=1}^{\infty} \frac{1 - e^{-k_n^2 T}}{N_n k_n^2} \left\{ -\frac{A \rho_R/\rho_L}{L + \frac{1}{B} + \frac{\rho_L}{A}} S_1 +\left(  \frac{A\bar{\rho}}{\rho_L} - \frac{\rho_R}{L + \frac{1}{B} + \frac{\rho_L}{A}} \right) S_2 \right\} \right] (e^{-\lambda} - 1).
  \end{split}
\end{align}
$S_1$ and $S_2$ are the parameters that are defined as:
\begin{subequations}
\begin{align}
  S_1&\equiv\int_0^LxX_n(x)dx=\frac{L\sin k_nL}{k_n}+\frac{\rho_L\cos k_nL-AL\cos k_nL-\rho_L}{\rho_Lk_n^2}+\frac{A\sin k_nL}{\rho_Lk_n^3},\\
  S_2&\equiv\int_0^LX_n(x)dx=\frac{\sin k_nL}{k_n}+\frac{A(1-\cos k_nL)}{\rho_Lk_n^2}.
\end{align}
\end{subequations}
This result implies that the statistics of the integrated current are given by the difference between two independent Poisson processes (i.e., the Skellam distribution).
As shown in Appendix $\ref{AppendixC}$, this result is consistent with the microscopic derivation based on the motion of individual Brownian particles.

\subsubsection*{Quenched initial condition}
Under the quenched initial condition, the governing MFT equations remain identical to those for the annealed case $\eqref{eq;MFTRBM}$, but the initial density profile $\rho(x,0)$ is deterministically fixed. 
While the spatial boundary conditions remain the same as in the annealed initial condition, the temporal boundary conditions are modified as follows:
  \begin{align}
    H(x,0)=\text{not fixed}, \qquad H(x,T)=0.
\end{align}
In the same way as annealed initial condition, this MFT equations and boundary conditions can be linearized by using Cole-Hopf transformation $P=e^H$, $Q=\rho e^{-H}$. The spatial boundary conditions remain the same as those for the annealed initial condition$\eqref{eq;spaceboundary}$, while the temporal boundary conditions are modified as follows:
\begin{align}
  P(x,T)=1, \qquad Q(x,0)=\frac{\bar{\rho}}{P(x,0)}.
\end{align}
Solving these equations, $P(x,t)$ and $Q(x,t)$ are obtained as
\begin{align}\label{eq;pqanneal}
  P(x,t)&=a_px+b_p+\int_0^LG^R(x,y;T-t)(1-a_py-b_p)dy,\\
  Q(x,t)&=a_qx+b_q+\int_0^LG^R(x,y;t)\qty(\frac{\bar{\rho}}{P(y,0)}-a_qy-b_q).
\end{align}
$a_p,\ b_p,\ a_q$ and $b_q$ are defined in $\eqref{eq;a_p}$. From $\eqref{eq;pqanneal}$ and $\eqref{eq;CGFderivative}$, we obtain current SCGF for quenched initial condition
\begin{align}
  \mu_{\mathcal{Q}}(\lambda)=\mu_{\mathcal{A}}(\lambda)+\int_0^{\lambda}d\lambda'\int_0^Tdt\int_0^Ldy\qty(a_p-\frac{A}{\rho_L} b_p)\qty(\frac{\bar{\rho}}{P(y,0)}-\bar{\rho})G^R(0,y;t)
\end{align}

\subsubsection*{Comparison of cumulants}

 We discuss the influence of initial fluctuations on current fluctuations under relaxation. We compare the current cumulants of annealed initial condition and quenched initial condition.  \\
From the exact solution for annealed and quenched initial condition, we could derive first and second cumulant of integrated current.
first cumulant is
\begin{align}
  \ev{Q_T}_{c,Q}=\ev{Q_T}_{c,A}
\end{align}
Current variance has the relation
\begin{align}
  \ev{Q_T^2}_{c,Q}=\ev{Q_T^2}_{c,A}+\int_0^Tdt\int_0^Ldy C\partial_{\lambda}P(y,0)G^R(0,y,t)\left.\right|_{\lambda=0}
\end{align}
From $C>0,\ G^R(x,y,t)>0$, and $\partial_{\lambda}P(y,0)\left.\right|_{\lambda=0}<0$, the second term is negative; hence, we obtain
\begin{align}
  \ev{Q_T^2}_{c,Q}<\ev{Q_T^2}_{c,A}
\end{align}

\subsubsection*{Large deviation function}
From the Legendre transformation of $\eqref{cumulant MFT}$, the large deviation function of the current under the relaxation process is obtained. 
\begin{figure}[H]
  \centering
  \begin{minipage}[c]{0.55\textwidth}
    \centering
    \includegraphics[width=\textwidth]{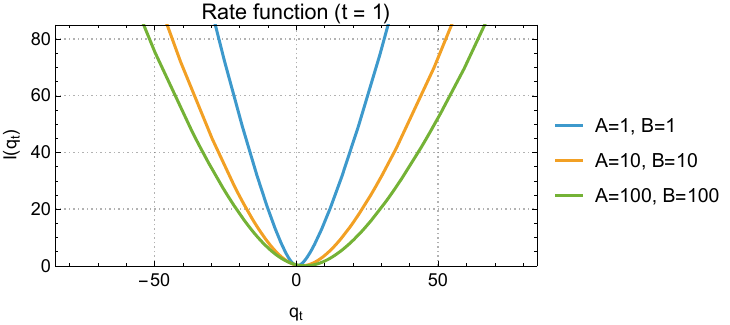}
  \end{minipage}
  \hfill 
  \begin{minipage}[c]{0.4\textwidth}
    \caption{Large deviation function $I(q)$ for the annealed initial condition, obtained via the Legendre transformation of the SCGF in $\eqref{cumulant MFT}$. The plots illustrate the dependence on the boundary coupling strengths $(A,B)$, comparing the case $(1,1),\ (10,10)$ and $(100,100)$. Other parameters are set as $L=5,\ \rho_L=0.8, \ \rho_R=0.2, \ \brho=0.5$. (at $t=1$)}
    \label{fig:large deviation function(anneal)}
    \vspace{10pt}
  \end{minipage}
\end{figure}
\noindent{}As shown, the large deviation function becomes steeper as the boundary dynamics become slower. This is likely due to the fact that the slower boundaries lead to a lower particle injection per unit time.\\

We examine the current large deviation function for the marginal-boundary case ($A=B=10$), focusing on the transition between different scaling regimes.   
\begin{figure}[H]
  \centering
  \begin{minipage}[c]{0.49\textwidth}
    \centering
    \includegraphics[width=\textwidth]{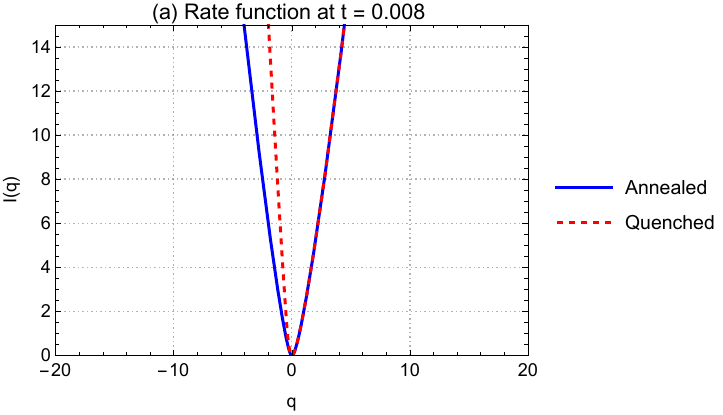}
  \end{minipage}
  \hfill 
   \begin{minipage}[c]{0.49\textwidth}
    \centering
    \includegraphics[width=\textwidth]{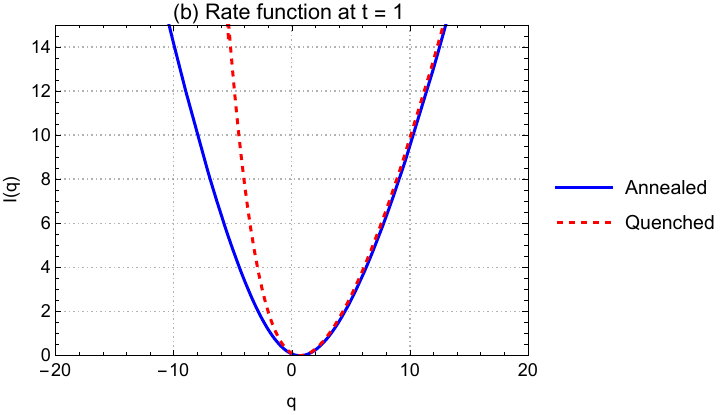}
  \end{minipage}
\end{figure}

\begin{figure}[H]
  \centering
  \begin{minipage}[c]{0.49\textwidth}
    \centering
    \includegraphics[width=\textwidth]{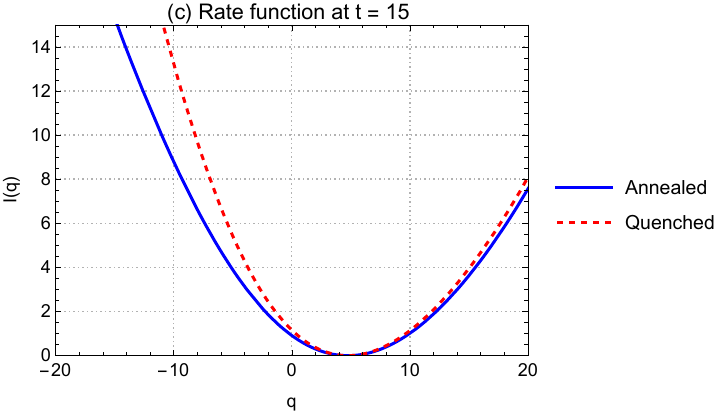}
  \end{minipage}
  \hfill 
  \begin{minipage}[c]{0.49\textwidth}
    \caption{The plots compare the LDFs under annealed and quenched initial conditions across three distinct scaling regimes: (a)the initial $O(T)$ regime at $t=0.008$,  (b)the intermediate diffusive $O(\sqrt{T})$ regime at $t=1$, (c)$O(T)$ steady-state regime at $t=15$.   Under the parameters $L=5,\ \rho_L=2, \ \rho_R=0.5, \ \brho=1.25$.}
    \label{fig:large deviation function(anneal)}
    \vspace{10pt}
  \end{minipage}
\end{figure}
In the negative current regime $(q<0)$, the large deviation function of quenched case exhibits a more rapid increase compared to the annealed case, implying that negative current fluctuations are less likely to occur when the initial configuration is fixed.

\subsection{Result for fast boundary}\label{sec;fast}
To obtain a simpler expression and to compare with the results of previous studies, we mention the result for fast-boundary. In the case of fast-boundary, the spatial boundary conditions are given by
\begin{align}
  H(0,t)=-\lambda, \qquad \rho(0,t)=\rho_L,\qquad
  H(L,t)=0, \qquad \rho(L,t)=\rho_R,
\end{align}
and temporal boundary conditions are
\begin{subequations}
\begin{align}
  \text{Annealed initial condition} \quad H(x,0)&=\log\rho-\log\bar{\rho},\qquad H(x,T)=0,\\
  \text{Quenched initial condition} \quad H(x,T)&=0.
\end{align}
\end{subequations}
In this  case, current SCGF for annealed initial condition is given by
\begin{align}\label{annealcgf}
\begin{split}
  \mu_{A}(\lambda)
  &=\frac{\rho_L(e^{\lambda}-1)+\rho_R(e^{-\lambda}-1)}{L}T\\ 
  &\qquad+L\sum\limits_{n=1}^{\infty}\frac{2\rho_L(e^{\lambda}-1)+2(e^{-\lambda}-1)\{\rho_R(-1)^n+\bar{\rho}(1-(-1)^n)\}}{(n\pi)^2}\qty{1-e^{-\qty(\frac{n\pi}{L})^2T}},
  \end{split}
\end{align}
and quenched initial condition is given by 
\begin{align}
\begin{split}
  \mu_Q(\lambda)&=\frac{\rho_L(e^{\lambda}-1)+\rho_R(e^{-\lambda}-1)}{L}T\\
  &\qquad+L\sum\limits_{n=1}^{\infty}\frac{2\rho_L(e^{\lambda}-1)+2(e^{-\lambda}-1)\rho_R(-1)^n-n\pi q_n}{(n\pi)^2}\qty{1-e^{-\qty(\frac{n\pi}{L})^2T}}. \label{eq;cgf(quench)}
\end{split}
\end{align}
$q_n$ is the parameter that is defined as follows:
\begin{align}
  q_n=\frac{2}{L}\int_0^{\lambda}d\lambda'\int_0^Ldx\frac{\bar{\rho}e^{-\lambda'}}{P(x,0)}\sin(\frac{n\pi}{L}x)
\end{align}
The $P(x,0)$ is the $\eqref{eq;pqanneal}$.

\subsubsection*{Large deviation function}
This is the time evolution of the large deviation function for annealed initial condition.
\begin{figure}[H]
  \centering
  \begin{minipage}[c]{0.55\textwidth}
    \centering
    \includegraphics[width=\textwidth]{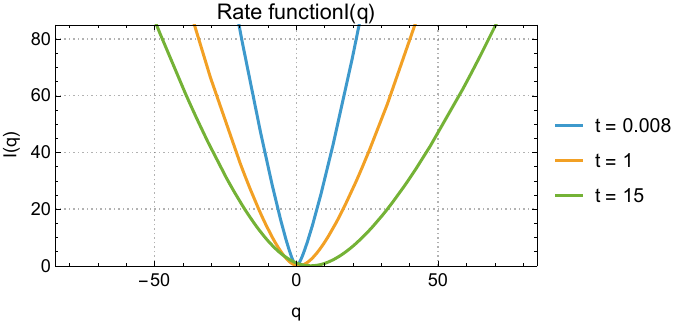}
  \end{minipage}
  \hfill 
  \begin{minipage}[c]{0.4\textwidth}
    \caption{Calculated results for the large deviation function from Legendre transformation of $\eqref{cumulant MFT}$  with parameters $L=5,\ \rho_L=2.0, \ \rho_R=0.5, \ \brho=1.25$. (at $t=0.008,1,15$)}
    \label{fig:large deviation function(anneal)}
    \vspace{10pt}
  \end{minipage}
\end{figure}

\subsubsection*{Consistency with previous research}
In this section, we verify whether the current SCGF for the finite non-stationary system obtained above is consistent with previous results for semi-infinite systems and the NESS. Although existing results were derived for IRW, we compare them by interpreting the IRW results as those for RBM, owing to the statistical equivalence between the two models.

The current fluctuations of the system in the limit $L\rightarrow\infty$ should correspond to those of the semi-infinite line system.
We now check the consistency between $L\rightarrow\infty$ and the result of semi-infinit system in the case of annealed initial condition. Taking $L\rightarrow\infty$ of equation $\eqref{annealcgf}$, the first term obviously vanishes. As for the second term, we will evaluate it after expressing it as an integral.
Separating the terms into oscillating and non-oscillating parts (We set $l_n=n\pi/L$),
\begin{equation*}
\begin{split}
  \text{(\text{the 2nd term})}&=\lim\limits_{L\rightarrow\infty}\sum\limits_{n=1}^{\infty}\frac{2\rho_L(e^{\lambda}-1)+2\bar{\rho}(e^{-\lambda}-1)+2(e^{-\lambda}-1)(\rho_R-\bar{\rho})(-1)^n}{(n\pi)^2}\qty{1-e^{-l_n^2T}}L\\
  &=\frac{1}{\pi}\lim\limits_{L\rightarrow\infty}\frac{\pi}{L}\sum\limits_{n=1}^{\infty}\qty{2\rho_L(e^{\lambda}-1)+2\bar{\rho}(e^{-\lambda}-1)+2(e^{-\lambda}-1)(\rho_R-\bar{\rho})\cos{l_nL}}\frac{1-e^{-l_n^2T}}{l_n^2}\\
  &=\frac{2\rho_L(e^{\lambda}-1)+2\bar{\rho}(e^{-\lambda}-1)}{\pi}\int_0^{\infty}\frac{1-e^{-k^2T}}{k^2}dk+\frac{2(e^{-\lambda}-1)(\rho_R-\bar{\rho})}{\pi}\lim\limits_{L\rightarrow\infty}\int_0^{\infty}dk\cos{kL}\frac{1-e^{-k^2T}}{k^2}
\end{split}
\end{equation*}
Here, the second term on the far right-hand side vanishes by the Riemann–Lebesgue Lemma. By calculating the integral in first term, we obtain the asymptotic form of $\mu(\lambda)$ as $L\rightarrow\infty$:
\begin{align}
\lim\limits_{L\rightarrow\infty}\mu(\lambda)=\frac{2\rho_L(e^{\lambda}-1)+2\bar{\rho}(e^{-\lambda}-1)}{\sqrt{\pi}}\sqrt{T}
\end{align}
This result is consistent with that of the semi-infinite system under the annealed initial condition \cite{Saha2023}.\\

Similarly, the current fluctuations of the system in the long-time average should correspond to those of the steady state current. Considering long-time average of $\eqref{annealcgf}$, we get SCGF for steady state current per unit time as follows: 
\begin{align}\label{eq;ness}
  \mu_{NESS}(\lambda)=\frac{\rho_L(e^{\lambda}-1)+\rho_R(e^{-\lambda}-1)}{L}
\end{align}
The SCGF of the current for IRW in a NESS is obtained.

\subsection*{First and second cumulants}
We can derive explicit formula of first and second cumulants.
The first cumulant is the same for both annealed system and quenched system, 
 \begin{align}
   \ev{Q_T}_{c,\mA}=\ev{Q_T}_{c,\mQ}=\frac{\rho_L-\rho_R}{L}T+L\sum\limits_{n=1}^{\infty}\frac{2\rho_L-2\{\rho_R(-1)^n+\brho(1-(-1)^n)\}}{(n\pi)^2}\qty{1-e^{-\qty(\frac{n\pi}{L})^2T}} 
 \end{align}
 Initial fluctuations do not affect mean dynamics. However, the second cumulant is different for annealed system and quenched system,
 \begin{subequations}
 \begin{align}
   \ev{Q_T^2}_{c,\mA}&=\frac{\rho_L+\rho_R}{L}T+L\sum\limits_{n=1}^{\infty}\frac{2\rho_L+2\{\rho_R(-1)^n+\brho(1-(-1)^n)\}}{(n\pi)^2}\qty{1-e^{-\qty(\frac{n\pi}{L})^2T}},\\
   \ev{Q_T^2}_{c,\mQ}&=\frac{\rho_L+\rho_R}{L}T+L\sum\limits_{n=1}^{\infty}\frac{2\rho_L+2\{\rho_R(-1)^n+\brho(e^{-\qty(\frac{n\pi}{L})^2t}-(-1)^n)\}}{(n\pi)^2}\qty{1-e^{-\qty(\frac{n\pi}{L})^2T}}.
 \end{align}
 \end{subequations}
From these relations, we can also check the inequality $\ev{Q_T^2}_{c,\mA}> \ev{Q_T^2}_{c,\mQ}$.

\section{Conclusion and Outlook}

In this study, we present an exact analysis of non-steady current fluctuations during the relaxation process of a finite system coupled to particle reservoirs at both ends—a regime that has so far remained unexplored within the MFT framework. Specifically, for stochastic lattice gases, we derived the MFT equations and their associated boundary conditions by taking the continuum limit of the moment-generating function for the integrated current. Through this approach, we achieved the following two major results:
\begin{itemize}
  \item Exact derivation of current variance for diffusive systems: For systems characterized by a constant diffusion coefficient D and arbitrary mobility $\sigma(\rho)$, we derived the exact analytical expression for the current variance. By evaluating both annealed and quenched initial conditions, we quantitatively assessed the impact of initial fluctuations on the current fluctuations during the relaxation process.
  \item Cumulant generating function for RBM: We derived the cumulant generating function of the integrated current for RBM with arbitrary boundary speeds.
\end{itemize}
We confirmed that our results are consistent with previous research. These findings demonstrate that MFT serves as a powerful framework for describing non-equilibrium fluctuations, not only for non-equilibrium steady states but also for the time-dependent processes of finite systems where boundaries and initial conditions play a dominant role.\\

One of the key future directions is to determine the large deviations of other time-dependent physical quantities, such as the density field, in addition to the current. Such an analysis would not only clarify the spatio-temporal structure of fluctuations but also help us investigate their universal properties during the relaxation process toward a non-equilibrium steady state. This, in turn, could lead to a better understanding of the thermodynamic properties of relaxation processes.

Although our exact evaluation of the current SCGF is specific to the RBM, extending the derivation of the full SCGF to systems with strong many-body interactions, such as the symmetric simple exclusion process (SEP), remains an important challenge. Notably, the MFT equations for the infinite-system SEP have been solved exactly by mapping them to classical integrable systems via non-local transformations\cite{Mallick2022}. Solving the MFT equations for a finite-system SEP through a similar mapping to integrable systems is an important task at the intersection of non-equilibrium physics and integrability system.

\section*{Acknowledgements}
DS and TS are grateful to Kazuya Fujimoto and Gaku Ohta for helpful discussions and comments. We are also grateful to Timur Aslyamov for his comments, which suggested the relation between our results and \cite{aslyamov_dynamical_2026}. This advice led to simpler expressions for the current variances in Section 3. The work of TS has been supported by JSPS KAKENHI Grants No. JP21H04432, No. JP22H01143, and No. JP23K22414.

\bibliographystyle{ieeetr}
\bibliography{references20260619}  

\appendix
\crefalias{section}{appendix}

\section{Derivation of $\eqref{eq:Drho}$ and $\eqref{eq:sigmarho}$ from $\eqref{eq:D_gradient}$ and $\eqref{eq:sigma_gradient}$}\label[appendix]{app:A1}

We show the consistency between the two sets of definitions, $\eqref{eq:Drho}$,$\eqref{eq:sigmarho}$ and $\eqref{eq:D_gradient}$ and $\eqref{eq:sigma_gradient}$. Specifically, we demonstrate that $\eqref{eq:Drho}$ and $\eqref{eq:sigmarho}$ can be derived from the definitions $\eqref{eq:D_gradient}$ and $\eqref{eq:sigma_gradient}$.
Let the integrated current $Q_T$ be defined as $Q_T=\frac{1}{L}\sum\limits_{i=1}^{L}\sum\limits_{k=0}^{M-1}Y_i(kd\tau)$, where $Y_i(kd\tau)$ is the stochastic variable defined in \eqref{eq:Y_i}.

\paragraph{Diffusion coefficient $D(\rho)$:}
Taking the expectation value of $Q_T$ with respect to $\{Y_i(kd\tau)\}$, we obtain
\begin{align}
  \ev{Q_T}=\frac{1}{L}\sum_{i=1}^{L}\sum_{k=0}^{M-1} (c^{i,i+1}-c^{i+1,i}) d\tau.
\end{align}
By assuming the gradient condition, adjacent terms in the spatial summation cancel out, leaving only the boundary terms:
\begin{align}
  \ev{Q_T}=\frac{1}{L}\sum_{k=0}^{M-1} \qty[ h(\tau_1\eta) - h(\tau_{L+1}\eta) ] d\tau.
\end{align}
Under the local equilibrium approximation for a sufficiently large lattice ($L \gg 1$), the local function $h(\tau_i\eta)$ can be effectively replaced by its macroscopic expectation value $\Phi(\rho(x))$. In the continuous time limit $d\tau \to 0$, the summation over time steps reduces to a time integral, yielding
\begin{align}
  \ev{Q_T}&\simeq\frac{1}{L}\int_0^T dt \qty[ \Phi(\rho_L) - \Phi(\rho_R) ] \nonumber \\
  &\simeq \frac{1}{L}\int_0^T dt \, D(\rho)\Delta\rho,
\end{align}
where $\Delta\rho = \rho_L - \rho_R$. We used $\Phi(\rho_L)-\Phi(\rho_R)\simeq \Phi'(\rho)\Delta\rho$ and the definition $D(\rho)\equiv \Phi'(\rho)$. Thus, the relation $\lim\limits_{T\rightarrow\infty}\frac{\ev{Q_T}}{T}=\frac{D(\rho)}{L}\Delta\rho$ holds.

\paragraph{Mobility $\sigma(\rho)$:}
Next, we evaluate the second moment $\ev{Q_T^2}$:
\begin{align}
  \ev{Q_T^2}=\frac{1}{L^2}\sum_{i,j=1}^{L}\sum_{k,l=0}^{M-1}\ev{Y_i(kd\tau)Y_j(ld\tau)}.
\end{align}
Under the assumption of $\rho_L = \rho_R = \rho$ in $\eqref{eq:sigmarho}$ (we assume that the system is in equilibrium), we have $\langle Y_i \rangle = 0$. 
Due to the Markov property, the full history up to the past time step does not bias the future fluctuation, ensuring that the cross-time correlations vanish exactly: $\ev{Y_i(kd\tau)Y_j(ld\tau)}=0$ for $k \neq l$.
Furthermore, the equal-time cross-correlations between different sites ($i \neq j$) scale as $\sum_{k=0}^{M-1}\ev{Y_i(kd\tau)Y_j(kd\tau)}=\sum_{k=0}^{M-1}O(d\tau^2)=O(d\tau)$, which vanishes in the continuous time limit $d\tau \to 0$.
Therefore, only the diagonal terms survive:
\begin{align}
  \ev{Q_T^2}&=\frac{1}{L^2}\sum\limits_{i=1}^{L}\sum\limits_{k=0}^{M-1}\ev{Y_i^2(kd\tau)} \nonumber \\
  &=\frac{1}{L^2}\sum\limits_{i=1}^{L}\sum\limits_{k=0}^{M-1} (c^{i,i+1}+c^{i+1,i}) d\tau.
\end{align}
Applying the local equilibrium assumption, the sum of the transition rates $c^{i,i+1}+c^{i+1,i}$ is effectively replaced by the mobility $\sigma(\rho)$ defined in \eqref{eq:sigma_gradient}. In the continuous time limit $d\tau \to 0$, since $\sigma(\rho)$ does not depend on $x$ and $t$, this yields
\begin{align}
  \ev{Q_T^2}=\frac{1}{L^2} \sum_{i=1}^L \int_0^T dt \, \sigma(\rho) = \frac{T}{L}\sigma(\rho).
\end{align}
Thus, we successfully retrieve the $\eqref{eq:sigmarho}$: $\lim\limits_{T\rightarrow\infty}\frac{\ev{Q_T^2}}{T}=\frac{\sigma(\rho)}{L}$.

\section{ Derivation of MFT action}\label{ap;MFT action}
In this appendix, we derive the MFT action for a boundary-driven diffusive system. We assume the stochastic lattice gases that particles hopping on a one dimensional lattice as a continuous-time Markov jump process $\eqref{eq;Markovgenerator}$.
The derivation of the MFT action in this section adapts the method used by \cite{Lefevre2007,Saha2023,Saha2024} for the SEP in semi-infinite systems and in the NESS to the case of stochastic lattice gases in finite and non-stationary systems.

Number of particles moving from site $i$ to $i+1$ in the infinitesimal time interval $[\tau,\tau+d\tau]$ is denoted by $Y_i(\tau)$. In the bulk ($2\leq i\leq N-1$) ,
\begin{align}\label{eq:Y_i}
\begin{split}
  Y_i(\tau)=\begin{cases}
    1\qquad &\text{with prob} \qquad c^{i,i+1}(\eta)d\tau\\
    -1\qquad &\text{with prob}\qquad c^{i+1,i}(\eta)d\tau\\
    0 \qquad &\text{with prob}\qquad 1-[c^{i,i+1}(\eta)+c^{i+1,i}(\eta)]d\tau
  \end{cases}
  \end{split}
\end{align}
Particles are injected and removed at the boundaries. We identify particles entering from the reservoir as being created and those exiting as being annihilated. The number of creation and annihilation events $Y_1^{\pm}$ in the infinitesimal time interval $[\tau, \tau+d\tau]$ can be expressed using the rates $c^{\pm}_1$ as follows:
\begin{align}
\begin{split}
  Y_1^{\pm}(\tau)=\begin{cases}
    1\qquad &\text{with prob}\qquad c_1^{+}(\eta) d\tau\\
    -1\qquad &\text{with prob}\qquad c_1^-(\eta) d\tau\\
    0 \qquad &\text{with prob}\qquad 1-[c_1^{+}(\eta)+c_1^-(\eta)]d\tau
  \end{cases}
  \end{split}
\end{align}
Similarly, for the right reservoir,
\begin{align}
\begin{split}
   Y_{N}^{\pm}(\tau)=\begin{cases}
    1\qquad &\text{with prob}\qquad c_N^+(\eta)d\tau\\
    -1\qquad &\text{with prob}\qquad c_N^-(\eta) d\tau\\
    0 \qquad &\text{with prob}\qquad 1-[c_N^+(\eta)+c_N^-(\eta)]d\tau
  \end{cases}
  \end{split}
\end{align}
In terms of $Y_i$, $Y_1^{\pm}$, and $Y_N^{\pm}$, the moment generating function of the current $\ev{e^{\lambda Q_T}}$ admits the following path integral representation:
\begin{align} \label{discretemoment}
  \ev{e^{\lambda Q_T}} = \int \mathcal{D}[n,Y] \bigg\langle 
  &\prod_{k=0}^{M-1} e^{\lambda Y_1^{\pm}(kd\tau)} \delta_{n_1(kd\tau+d\tau)-n_1(kd\tau), Y_1^{\pm}(kd\tau)-Y_1(kd\tau)} \notag \\
  &\times \prod_{i=2}^{N-1} \delta_{n_i(kd\tau+d\tau)-n_i(kd\tau), Y_{i-1}(kd\tau)-Y_i(kd\tau)} \notag \\
  &\times \delta_{n_N(kd\tau+d\tau)-n_N(kd\tau), Y_N^{\pm}(kd\tau)-Y_{N-1}(kd\tau)} \bigg\rangle_{[Y]}.
\end{align}
We divide the time interval $[0,T]$ into $M$ sub-intervals of equal length $d\tau=T/M$. The Kronecker delta expresses the law of particle conservation. The path integral measure on $n$ and $Y$ are defined as
\begin{align}
  \int\mathcal{D}[n]=\prod_{k=1}^{M}\prod_{i=1}^{N}\sum\limits_{n_i(kd\tau)=0}^{n},\qquad \int\mathcal{D}[Y]=\prod_{k=0}^{M-1}\prod_{i=1}^{N-1}\sum\limits_{Y_i(kd\tau)=-1}^{1}\sum_{Y_1^{\pm}(kd\tau)=-1}^1\sum_{Y_N^{\pm}=-1}^1.
\end{align}
 $\ev{}_{[Y]}$ represents the expectation values by $(Y_1^{\pm}(kd\tau),Y_1(kd\tau),\cdots,Y_{N-1}(kd\tau),Y_N^{\pm}(kd\tau))$. By introducing the auxiliary field $\hat{n}_i$ and using the integral representation of the Kronecker delta,
$\delta_{a,b}=\frac{1}{2\pi i}\int_{-i\pi}^{i\pi}d\hat{n}_ie^{-\hat{n}_i(a-b)},\quad a,b\in\mathbb{Z}$, we obtain the following expression from $\eqref{discretemoment}$: 
\begin{align}\label{discretemoment2}
\begin{split}
  \ev{e^{\lambda Q_T}} &= \int\mathcal{D}[n,\hn,Y]P(\{n_i(0)\}) e^{-\sum\limits_{k=0}^{M-1}\sum\limits_{i=1}^{N}\hn_i(kd\tau)(n_i(kd\tau+d\tau)-n_i(kd\tau))} \\[1.5ex]
  &\quad \times\ev{e^{\lambda\sum\limits_{k=0}^{M-1}Y_1^{\pm}(kd\tau)+\sum\limits_{k=0}^{M-1}\hn_1(kd\tau)Y_1^{\pm}}}_{Y_1^{\pm}} \\[1.5ex]
  &\quad \times\ev{e^{-\sum\limits_{k=0}^{M-1}\hn_1Y_1(kd\tau)+\sum\limits_{k=0}^{M-1}\sum\limits_{i=2}^{N-1}\hn_i(kd\tau)(Y_{i-1}(kd\tau)-Y_i(kd\tau))+\sum\limits_{k=0}^{M-1}\hn_N(kd\tau)Y_{N-1}(kd\tau)}}_{[Y_i]} \\[1.5ex]
  &\quad \times \ev{e^{\sum\limits_{k=0}^{M-1}\hn_N(kd\tau)Y^{\pm}_{N}(kd\tau)}}_{Y_N^{\pm}}
\end{split}
\end{align}
$P(\{n_i(0)\})$ denotes the probability distribution of the initial fluctuations. 
Subsequently, we evaluate the expectation of each term. The left boundary term is
\begin{align}\label{left}
  \ev{e^{\lambda Y_1^{\pm}(kd\tau)+\hn_1(kd\tau)Y_1^{\pm}}}_{Y_1^{\pm}}&=1+\left[c_1^+\left(e^{\lambda+\hat{n}_1(kd\tau)}-1\right)+c_1^-\left(e^{-(\lambda+\hat{n}_1(kd\tau))}-1\right) \right]d\tau \notag\\
&\simeq\exp\left[c_1^+d\tau\left(e^{\lambda+\hat{n}_1(kd\tau)}-1\right)+c_1^-\left(e^{-(\lambda+\hat{n}_1(kd\tau))}-1\right) \right].
\end{align}
Similarly, right boundary term is
\begin{align}\label{right}
  \ev{e^{\hn_N(kd\tau)Y^{\pm}_{N}(kd\tau)}}_{Y_N^{\pm}}\simeq\exp\left[c_N^- d\tau \left(e^{-\hat{n}_{N-1}}-1 \right)+c_N^+ d\tau\left(e^{\hat{n}_{N-1}}-1  \right) \right],
\end{align}
and bulk term is
\begin{align}\label{bulk}
   &\ev{e^{-\sum\limits_{k=0}^{M-1}\hn_1(kd\tau)Y_1(kd\tau)+\sum\limits_{k=0}^{M-1}\sum\limits_{i=2}^{N-1}\hn_i(kd\tau)(Y_{i-1}(kd\tau)-Y_i(kd\tau))+\sum\limits_{k=0}^{M-1}\hn_N(kd\tau)Y_{N-1}(kd\tau)}}_{[Y_i]}\notag\\
   &\simeq\prod_{k=0}^{M-1}\ev{\exp\qty(\sum\limits_{i=1}^{N-1}(\hn_{i+1}-\hn_i)Y_i)}_{[Y_i]}\notag\\
   &=\prod_{k=0}^{M-1}\prod_{i=1}^{N-1}\qty[\exp\qty{c^{i,i+1}(e^{\hn_{i+1}-\hn_i}-1)d\tau+c^{i+1,i}(e^{-(\hn_{i+1}-\hn_i)}-1)d\tau }].
\end{align}
Substituting $\eqref{discretemoment2}$ into $\eqref{left}$, $\eqref{right}$, $\eqref{bulk}$ and taking the continuous-time limit, we obtain
\begin{align}\label{discretemoment3}
\begin{split}
  \ev{e^{\lambda Q_T}} &= \int\mathcal{D}[n,\hn,Y]P(\{n_i(0)\})\exp\qty[\underbrace{-\int_0^{T}dt\sum\limits_{i=1}^{N}\hn_i(t)\dv{n_i(t)}{t}}_{(1)}] \\[1.5ex]
  &\quad \times\exp\qty[\underbrace{\int_0^{T}dt\qty{c_1^+\left(e^{\lambda+\hat{n}_1}-1\right)+c_1^-\left(e^{-(\lambda+\hat{n}_1)}-1\right)}}_{(L)} ] \\[1.5ex]
  &\quad \times\exp \qty[\underbrace{\int_0^{T}dt\sum_{i=1}^{N-1}\qty{c^{i,i+1}(e^{\hn_{i+1}-\hn_i}-1)+c^{i+1,i}(e^{-(\hn_{i+1}-\hn_i)}-1) }}_{(bulk)}]\\
  &\quad \times\exp\qty[\underbrace{\int_0^{T}dt\qty{c_N^- \left(e^{-\hat{n}_{N-1}}-1 \right)+c_N^+ \left(e^{\hat{n}_{N-1}}-1  \right)}}_{(R)}]
\end{split}
\end{align}
In the following, following $\eqref{eq;diffusivescale}$, we take diffusive scaling limit $\Lambda\rightarrow\infty$ and investigate the asymptotic representation of $\eqref{discretemoment3}$. In the following, we set $\Delta x=1/\Lambda$. By assuming $\hat{n}_i(t)\rightarrow H(x,t)$, $n_i(t)\rightarrow \rho(x,t)$ for $\Delta x\rightarrow0$,
\begin{align}
  (1)&=-\int_0^{T}dt\sum\limits_{i=1}^{N}\hn_i(t)\dv{n_i(t)}{t}\notag\\
  &\stackrel{\mathclap{\Lambda\rightarrow\infty}}{\simeq}\;\;-\Lambda\int_0^Tdt\int_0^LdxH(x,t)\pdv{\rho(x,t)}{t}.
\end{align}
The symbol $\stackrel{\Lambda\rightarrow\infty}{\simeq}$ denotes the asymptotic behavior in the limit $\Lambda\rightarrow\infty$.
In the same way, derive the asymptotic representation for the left boundary. When the system scale is transformed as $\eqref{eq;diffusivescale}$, the transition rate at the boundaries $c_1^{\pm}$ and $c_N^{\pm}$ scale as $c_1^{\pm}=C_1^{\pm}/\Lambda$ and $c_N^{\pm}=C_N^{\pm}/\Lambda$, respectively. 
\begin{align}
  (L)&=\Lambda^2\int_0^{T}dt\qty{\frac{C_1^+}{\Lambda}\left(e^{\lambda+\hat{n}_1(t)}-1\right)+\frac{C_1^-}{\Lambda}\left(e^{-(\lambda+\hat{n}_1(t))}-1\right)}\notag\\
  &\stackrel{\mathclap{\Lambda\rightarrow\infty}}{\simeq}\;\; \Lambda\int_0^Tdt\qty{C_1^+(\rho_L,\rho(0,t))\qty(e^{\lambda+H(0,t)}-1) +C_1^-(\rho_L,\rho(0,t))\qty(e^{-\lambda-H(0,t)}-1)}.
\end{align}
Similarly, asymptotic representation for the right boundary is
\begin{align}
  (R)&=\Lambda^2\int_0^{T}dt\qty{\frac{C_N^-}{\Lambda} \left(e^{-\hat{n}_{N-1}}-1 \right)+\frac{C_N^+}{\Lambda} \left(e^{\hat{n}_{N-1}}-1  \right)}\notag\\
  &\stackrel{\mathclap{\Lambda\rightarrow\infty}}{\simeq}\; \;\Lambda\int_0^Tdt\qty{C_N^-(\rho_R,\rho(L,t))\qty(e^{-H(L,t)}-1)+C_N^+(\rho_R,\rho(L,t))\qty(e^{H(L,t)}-1)}.
\end{align}
The bulk term is given by 
\begin{align}\label{eq;Hbulk}
  (bulk)=\Lambda^2\int_0^Tdt \underbrace{\sum_{i=1}^{\Lambda N-1}\qty{c^{i,i+1}(e^{\hn_{i+1}-\hn_i}-1)+c^{i+1,i}(e^{-(\hn_{i+1}-\hn_i)}-1) }}_{\text{Bulk}}
\end{align}
To proceed to the continuum limit, we expand up to the second order in $\Delta x$, consistent with the diffusive scaling $(\Delta x)^2\sim \Delta t$.\\
\begin{align}\label{eq:expandH}
  e^{\pm(\hn_{i+1}-\hn_i)}-1&=\pm\dv{\hn_i}{x}\Delta x+\frac{1}{2}\qty(\pm\dv[2]{\hn_i}{x}+\qty(\dv{\hn_i}{x})^2)(\Delta x)^2+O((\Delta x)^3)
\end{align}
By inserting $\eqref{eq:expandH}$ into $\eqref{eq;Hbulk}$, we obtain 
\begin{equation}
\label{eq:Bulk}
\begin{split}
 \text{(Bulk)}=&\sum_{i=1}^{\Lambda N-1}\Big[ \qty(c^{i,i+1}(\eta)-c^{i+1,i}(\eta))\dv{\hn_i}{x}\Delta x\\
 &\quad+\frac{1}{2}\Big\{c^{i,i+1}(\eta)\Big(\dv[2]{\hn_i}{x}+\qty(\dv{\hn_i}{x})^2\Big)+c^{i+1,i}(\eta)\Big(-\dv[2]{\hn_i}{x}+\qty(\dv{\hn_i}{x})^2\Big)\Big\}(\Delta x)^2+O((\Delta x)^3)\Big].
\end{split}
\end{equation}
To take the diffusive scaling limit, we introduce two assumptions \cite{Spohn1991,Kipnis1999,Bertini2006}. First, we assume a gradient-type system where the microscopic current is given by the difference of a local function $h(\eta)$ at adjacent sites. This gradient condition is assumed to ensure the tractability of the hydrodynamic limit. In this case, the condition $c^{i,i+1}(\eta)-c^{i+1,i}(\eta)=h(\tau_i\eta)-h(\tau_{i+1}\eta)$ is satisfied ($\tau_i$ is the shift operator that is defined as $[\tau_i\eta](j)=\eta(j-i)$). Second, we assume that local equilibrium holds everywhere in the system. Under the local equilibrium assumption, $h(\tau_i\eta)$ is replaced by expected value $\Phi(\rho(i\Delta x))\equiv \mathbb{E}_{P^{\rho}_{eq}}[h(\tau_i\eta)]$ under the equilibrium distribution at density $\rho$ near site $i$. The diffusion coefficient $D(\rho)$ is obtained as $\Phi'(\rho)=D(\rho)$. Also, mobility $\sigma(\rho)$ is obtained as  $\sigma(\rho)\equiv\mathbb{E}_{P_{eq}^{\rho}}[\qty(c^{i,i+1}(\eta)+c^{i+1,i}(\eta))]$.
The first term of $\eqref{eq:Bulk}$ is
\begin{align*}
  \sum_{i=1}^{\Lambda N-1} \qty(c^{i,i+1}(\eta)-c^{i+1,i}(\eta))\dv{\hn_i}{x}\Delta x
  &=\sum_{i=1}^{\Lambda N-1}\qty(h(\tau_i\eta)-h(\tau_{i+1}\eta))\dv{\hn_i}{x}\Delta x\\
  &\simeq \sum_{i=1}^{\Lambda N-1}\qty(\Phi(\rho(i\Delta x))-\Phi(\rho(i\Delta x+\Delta x)))\dv{\hn_i}{x}\Delta x \\
  &\stackrel{\mathclap{\Lambda\rightarrow\infty}}{\simeq}\;\; -\frac{1}{\Lambda}\int_0^Ldx\, \Phi'(\rho)\pdv{\rho}{x}\pdv{H}{x}.
\end{align*}
The second term of $\eqref{eq:Bulk}$ is
\begin{align*}
  \sum_{i=1}^{\Lambda N-1}\frac{1}{2}\qty(c^{i,i+1}(\eta)+c^{i+1,i}(\eta))\qty(\dv{\hn_i}{x})^2(\Delta x)^2+O((\Delta x)^3)\stackrel{\Lambda\rightarrow\infty}{\simeq} \frac{1}{\Lambda}\int_0^Ldx\, \frac{\sigma(\rho)}{2}\qty(\pdv{H}{x})^2\qquad 
\end{align*}
From above results, 
\begin{align}
  \ev{e^{\lambda Q_T}}\simeq e^{-\Lambda S[\rho,H]}
\end{align}
where,
\begin{align}
  S[\rho,H]=\mathcal{F}[\rho(x,0)]+\int_0^Tdt\qty[\int_0^LH(x,t)\pdv{\rho(x,t)}{t}dx-\qty(\mathcal{H}_{bdry}^{(L)}+\mathcal{H}_{bulk}+\mathcal{H}_{bdry}^{(R)})] 
\end{align}
with
\begin{subequations}
\begin{align}
  \mathcal{H}_{bdry}^L&=C_1^+(\rho(0,t),\rho_L)\left(e^{\lambda+H(0,t)}-1\right) +C_1^-(\rho(0,t),\rho_L)\left(e^{-\lambda-H(0,t)}-1\right)\\
  \mathcal{H}_{bulk}&=\int_0^Ldx \ \qty(\frac{\sigma(\rho)}{2}\partial_xH-D(\rho)\partial_x\rho)\partial_xH\\
  \mathcal{H}_{bdry}^R&=C_N^-(\rho(L,t),\rho_R)\left(e^{-H(L,t)}-1\right)+C_N^+(\rho(L,t),\rho_R)\left(e^{H(L,t)}-1\right)
\end{align}
\end{subequations}
$\mathcal{F}[\rho(x,0)]$ denotes initial fluctuations. In the case of uniform state density $\bar{\rho}$, 
\begin{align*}
  \mathcal{F}[\rho(x,0)]=\begin{cases}
    \int_0^Ldx\int_{\bar{\rho}(x)}^{\rho(x,0)}dr\frac{2D(\rho)[\rho(x,0)-r]}{\sigma(r)}\qquad &\text{Annealed initial condition}\\
    0 & \text{Quenched initial condition}
  \end{cases}
\end{align*}

\section{Perturbative calculation}\label{ap_gene}
In this appendix, we provide the details of the perturbative calculation performed in Section $\ref{general}$.
By substituting $\eqref{eq:p-lambda}$ into the MFT equations, it can be seen that $\rho_0$, $\rho_1$, $H_1$ satisfy the following equations.
\begin{subequations}
\begin{align}
  \partial_t\rho_0&=D\partial_x^2\rho_0,\\
  \partial_tH_1&=-D\partial_x^2H_1,\\
  \partial_t\rho_1&=D\partial_x^2\rho_1-\partial_x(\sigma(\rho_0)\partial_xH_1).
\end{align}
\end{subequations}

\subsubsection*{Annealed initial conditions}
Noting that $H(x,0)=f'(\rho)-f'(\bar{\rho})=\lambda f''(\bar{\rho})\rho_1+o(\lambda)$, the boundary conditions for $\rho_0$, $\rho_1$ and $H_1$ are given as follows:
\begin{subequations}
\begin{align}
  \rho_0(x,0)&=\bar{\rho}, \quad \rho_0(0, t)=\rho_L,\quad \rho_0(L, t)=\rho_R,\\
  H_1(x, T)&=0,\quad H_1(0,t)=-1, \quad H_1(L,t)=0,\\
  \rho_1(x,0)&=\frac{\sigma(\rho_0)}{2D}H_1(x,0),\quad \rho_1(0,t)=\rho_1(L,t)=0.
\end{align}
\end{subequations}
Since $\rho_0(x,t)$ and $H_1(x,t)$ follow the standard (anti) diffusion equations, they are given as follows:
\begin{align}\label{eq;rho_0}
  \rho_0(x,t)&=\frac{\rho_R-\rho_L}{L}x+\rho_L+\sum\limits_{n=1}^{\infty}\frac{2}{n\pi}\qty{-\rho_L+\rho_R(-1)^n+\bar{\rho}(1-(-1)^n)}e^{-D\qty(\frac{n\pi}{L})^2t}\sin\frac{n\pi}{L}x,\\ \label{eq;H1}
  H_1(x,t)&=\frac{x}{L}-1+\sum\limits_{n=1}^{\infty}\frac{2}{n\pi}e^{-D\qty(\frac{n\pi}{L})^2(T-t)}\sin\frac{n\pi}{L}x.
\end{align}
By using these $\rho_0$ and $H_1$, along with the Green's function for the diffusion equation($0<x<L$, with Dirichlet boundary conditions) \eqref{eq;GreenDirichlet}
\begin{align*}
  G^D(x,y;t)=\frac{2}{L}\sum\limits_{n=1}^{\infty}\sin\qty(\frac{n\pi}{L}x)\sin\qty(\frac{n\pi}{L}y)e^{-D\qty(\frac{n\pi}{L})^2t}
\end{align*}
$\rho_1(x,t)$ can be expressed as follows;
\begin{align}\label{eq:B8.6}
  \rho_1(x,t)=\frac{1}{2D}\int_0^LdyG^D(x,y;t)\sigma(\rho_0(y,0))H_1(y,0)+\int_0^tds\int_0^Ldy\partial_yG^D(x,y;t-s)(\sigma(\rho_0(y,s))\partial_y H_1(y,s))
\end{align}
Inserting the explicit forms of $\rho_0$, $H_1$ and $\rho_1$ into $\eqref{eq;variance}$, $\ev{Q_T^2}_c=\int_0^Tdt\qty{-D\partial_x\rho_1+\sigma(\rho_0)\partial_xH_1}\left.\right|_{x=0}$, we obtain $\eqref{eq;generalanneal}$:
\begin{align*}
\begin{split}
  \ev{Q_T^2}_{c,\mathcal{A}}&=-\frac{1}{2}\int_0^Tdt\int_0^LdyG_x(x=0,y;t)\sigma(\rho_0(y,0))H_1(y,0)\\
  &\qquad-\int_0^Tdt\int_0^Ldy\int_0^{t}dsDG_{xy}(x=0,y;t-s)(\sigma(\rho_0(y,s))\partial_y H_1(y,s))
  +\int_0^Tdt\sigma(\rho_0)\partial_xH_1\left.\right|_{x=0}.
\end{split}
\end{align*}
For the first term, by inserting explicit form of $G^D_x(0,y;t)$, 
\begin{align}
  (\text{1st term})&=-\frac{1}{2}\int_0^Tdt\int_0^Ldy\frac{2}{L}\sum\limits_{n=1}^{\infty}\frac{n\pi}{L}\sin\qty(\frac{n\pi}{L}y)e^{-D\qty(\frac{n\pi}{L})^2t}\sigma(\bar{\rho})H_1(y,0)\\
  &=\frac{1}{2D}\int_0^Ldy\sum\limits_{n=1}^{\infty}\frac{2\sin\qty(\frac{n\pi}{L}y)}{n\pi}\qty{e^{-D\qty(\frac{n\pi}{L})^2T}-1}\sigma(\bar{\rho})H_1(y,0)\\
  &=\int_0^Ldy\frac{\sigma(\bar{\rho})}{2D}(H_1(y,0))^2,
\end{align}
where we have used $H_1(y,0)=\sum\limits_{n=1}^{\infty}\frac{2\sin\qty(\frac{n\pi}{L}y)}{n\pi}\qty{e^{-D\qty(\frac{n\pi}{L})^2T}-1}$.
Next, we evaluate the remaining terms. The second term can be rearranged as:
\begin{align}\label{eq;gen_ap_quench2}
  (\text{2nd term})=-D\int_0^Ldy\int_0^Tds (\sigma(\rho_0(y,s))\partial_y H_1(y,s)) \int_s^{T}dtG^D_{xy}(x=0,y;t-s).
\end{align}
With the aid of the explicit representations of $G_{xy}^D$,
\begin{align}
  D\int_s^{T}dtG^D_{xy}(x=0,y;t-s)&=\frac{2}{L}\sum\limits_{n=1}^{\infty}\cos\qty(\frac{n\pi}{L}y)-\frac{2}{L}\sum\limits_{n=1}^{\infty}\cos\qty(\frac{n\pi}{L}y)e^{-D\qty(\frac{n\pi}{L})^2(T-s)}\\
  &=\delta(y)-\partial_yH_1(y,s).
\end{align}
By inserting this into $\eqref{eq;gen_ap_quench2}$,
\begin{align}\label{eq;2nd_term}
  (\text{2nd term})=\int_0^Ldy\int_0^Tds \sigma(\rho_0(y,s))(\partial_y H_1(y,s))^2- \int_0^Tdt\sigma(\rho_0)\partial_xH_1\left.\right|_{x=0}
\end{align}
Since the second term of $\eqref{eq;2nd_term}$ cancels out the third term of \eqref{eq;generalanneal}, the expression for the current variance simplifies to:
\begin{align}
  \ev{Q_T^2}_{c,\mathcal{A}}=\int_0^Ldy\int_0^Tdt\ \sigma(\rho_0(y,t))(\partial_y H_1(y,t))^2+\int_0^Ldy\frac{\sigma(\bar{\rho})}{2D}(H_1(y,0))^2
\end{align}

\subsubsection*{Quenched initial conditions}
In the case of quenched initial conditions, 
there is no constraint on $H_1(x,0)$; instead, the condition $\rho_1(x,0)=0$ is imposed. Consequently, the boundary conditions are given as follows;
\begin{subequations}
\begin{align}
  \rho_0(x,0)&=\bar{\rho}, \quad \rho_0(0, t)=\rho_L,\quad \rho_0(L, t)=\rho_R,\\
  H_1(x, T)&=0,\quad H_1(0,t)=-1, \quad H_1(L,t)=0,\\
  \rho_1(x,0)&=0,\quad \rho_1(0,t)=\rho_1(L,t)=0.
\end{align}
\end{subequations}
$\rho_0(x,t)$ and $H_1(x,t)$ are the same as $\eqref{eq;rho_0}$ and $\eqref{eq;H1}$, respectively. $\rho_1(x,t)$ is given as follows:
\begin{align}\label{eq:B8.8}
  \rho_1(x,t)=\int_0^tds\int_0^Ldy\partial_yG^D(x,y;t-s)(\sigma(\rho_0(y,s))\partial_y H_1(y,s))
\end{align}
Inserting the explicit forms of $\rho_0$, $H_1$ and $\rho_1$ into  $\eqref{eq;variance}$, $\ev{Q_T^2}_c=\int_0^Tdt\qty{-D\partial_x\rho_1+\sigma(\rho_0)\partial_xH_1}\left.\right|_{x=0}$, we obtain $\eqref{eq;generalquench}$:
\begin{align*}
  \ev{Q_T^2}_{c,\mathcal{Q}}=-\int_0^Tdt\int_0^Ldy\int_0^{t}dsDG^D_{xy}(x=0,y;t-s)(\sigma(\rho_0(y,s))\partial_y H_1(y,s))+\int_0^Tdt\sigma(\rho_0)\partial_xH_1\left.\right|_{x=0}
\end{align*}
Similarly to the annealed case, we obtain the following compact form:
\begin{align}
 \ev{Q_T^2}_{c,\mQ}=\int_0^Tdt\int_0^Ldy\, \sigma(\rho_0)(\partial_yH_1)^2.
\end{align}

\section{Current SCGF for semi-infinite line system}
\subsection{Derivation of current SCGF}\label{Appendix_si_cgf}
In this appendix, we derive current SCGF on a semi-infinite system with a constant diffusion coefficient and arbitrary general mobility, following the method presented in \cite{Krapivsky2012}. In the case of semi-infinite line, the integrated current $Q_T$ is represented as $Q_T=\int_0^{\infty}dy\qty{\rho(x,T)-\rho(x,0)}$. In this case, although MFT equation is the same as $\eqref{eq;MFTgeneral}$, boundary conditions are modified. In the following, we denote $\rho$, $H$ for semi-infinite system as $\rho^{si}$, $H^{si}$.

\subsubsection*{Quenched initial condition}
The boundary conditions are
\begin{align}
  \rho^{si}(0,t)=\rho_L, \quad H^{si}(0,t)=0, \quad \rho^{si}(x,0)=\bar{\rho},\quad H^{si}(x,T)=\lambda.
\end{align}
By conducting perturbative calculation, $\rho^{si}_0$ and $H^{si}_1$ are given as follows:
\begin{align}
  \rho^{si}_0(x,t)&=\rho_L+(\bar{\rho}-\rho_L)\erf\qty(\frac{x}{\sqrt{4Dt}}),\\
  H^{si}_1(x,t)&=\erf\qty(\frac{x}{\sqrt{4D(T-t)}}).
\end{align}
By setting $\rho_1(x,t)=-\partial_x\psi_{\mQ}$,   $\psi_{\mQ}(x,t)$ satisfies
\begin{align}\label{eq;psiQ}
  (\partial_t-D\partial_x^2)\psi_{\mQ}=\sigma(\rho^{si}_0)\partial_xH_1^{si}
\end{align}
with Neumann boundary condition $\partial_x\psi_{\mQ}(x=0,t)=\rho_1^{si}(0,t)=0$.\\
Current SCGF is given as follows\cite{Derrida2009a}:
\begin{align}
  \mu_{\mQ}(\lambda)=\lambda\int_0^{\infty}dx\qty{\rho_1^{si}(x,T)-\rho_1^{si}(x,0)}-\frac{1}{2}\int_0^Tdt\int_0^{\infty}dx\sigma(\rho_0^{si})(\partial_xH_1^{si})^2.
\end{align}
Performing a perturbative expansion at order $\lambda^2$, current variance is 
\begin{align}
  \ev{Q_T^2}_{c,\mQ}&=-2\int_0^{\infty}dx\qty{\rho_1^{si}(x,T)-\rho_1^{si}(x,0)}-\int_0^Tdt\int_0^{\infty}dx\sigma(\rho_0^{si})(\partial_xH_1^{si})^2\notag \\
  &=2\psi_{\mQ}(0,T)-\int_0^Tdt\int_0^{\infty}dx\sigma(\rho_0^{si})(\partial_xH_1^{si})^2.
\end{align}
 By solving $\eqref{eq;psiQ}$ under initial condition $\psi_{\mQ}(x,0)=0$, we obtain
$\psi_{\mQ}(0,T)=\int_0^Tdt\int_0^{\infty}dx\sigma(\rho_0^{si})(\partial_xH_1^{si})^2$. Therefore, the current variance is
\begin{align}
  \ev{Q_T^2}_{c,\mQ}=\int_0^Tdt\int_0^{\infty}dx\sigma(\rho_0^{si})(\partial_xH_1^{si})^2.
\end{align}

\subsubsection*{Annealed initial condition}
The boundary conditions are 
\begin{align}
   \rho^{si}(0,t)=\rho_L, \quad H^{si}(0,t)=0, \quad H^{si}(x,T)=\lambda,\quad H^{si}(x,0)=\lambda+\int_{\bar{\rho}}^{\rho(x,0)}\frac{2D}{\sigma(r)}dr.
\end{align}
By conducting perturbative calculation, $\rho_0$ and $H_1$ are given as follows:
\begin{align}
  \rho^{si}_0(x,t)&=\rho_L+(\bar{\rho}-\rho_L)\erf\qty(\frac{x}{\sqrt{4Dt}})\\
  H^{si}_1(x,t)&=\erf\qty(\frac{x}{\sqrt{4D(T-t)}}).
\end{align}
It is the same as quenched case. From this, initial condition of $\rho_1^{si}(x,t)$ is $\rho_1^{si}(x,0)=\cfrac{\sigma(\brho)}{2D}\qty{\erf\qty(\frac{x}{\sqrt{4DT}})-1}$. By setting $\rho^{si}_1(x,t)=-\partial_x\psi_{\mA}$, $\psi_{\mA}(x,t)$ satisfies
\begin{align}\label{eq;psiA}
  (\partial_t-D\partial_x^2)\psi_{\mA}=\sigma(\rho^{si}_0)\partial_xH_1^{si}
\end{align}
with Neumann boundary condition $\partial_x\psi_{\mA}(x=0,t)=\rho_1^{si}(0,t)=0$.\\
Current SCGF is given as follows\cite{Derrida2009a}:
\begin{align}
  \mu_{\mA}(\lambda)=-\int_0^{\infty}dx\int_{\brho}^{\rho(x,t)}dr\frac{2D}{\sigma(r)}(\rho(x,0)-r)+\mu_{\mQ}(\lambda).
\end{align}
Performing a perturbative expansion at order $O(\lambda^2)$, current variance is 
\begin{align}
  \ev{Q_T^2}_{c,\mA}=-\int_0^{\infty}dx\frac{2D(\rho_1(x,0))^2}{\sigma(\brho)}+2\psi_{\mA}(0,T)-\int_0^Tdt\int_0^{\infty}dx\sigma(\rho_0^{si})(\partial_xH_1^{si})^2.
\end{align}
From $\eqref{eq;psiA}$, 
\begin{align}
\psi_{\mA}(0,T)=\int_0^{\infty}dyG(x=0,y;T)\psi_{\mA}(y,0)+\int_0^Tds\int_0^{\infty}dyG(0,y;T-s)\sigma(\rho^{si}_0)\partial_xH_1^{si}
\end{align}
and
\begin{align*}
  \psi_{\mA}(x,0)=\frac{\sigma(\brho)}{2D}\qty[x\ \mathrm{erfc}\qty(\frac{x}{\sqrt{4DT}})+2\sqrt{\frac{DT}{\pi}}\qty(1-e^{-\frac{x^2}{4DT}})]
\end{align*}
$G(x,y;t)$ denotes Green's function of semi-infinite diffusion equation with Neumann boundary condition.
After some algebra,
\begin{align}
   \ev{Q_T^2}_{c,\mA}&=\int_0^Tdt\int_0^{\infty}dx\sigma(\rho_0^{si})(\partial_xH_1^{si})^2+\frac{(2-\sqrt{2})\sigma(\brho)\sqrt{T}}{\sqrt{D\pi}}\notag \\
   &=\ev{Q_T^2}_{c,\mQ}+\frac{(2-\sqrt{2})\sigma(\brho)\sqrt{T}}{\sqrt{D\pi}}.
\end{align}

\subsection{Consistency with our result}\label{Appendix_fi_cgf}
\subsubsection*{Quenched initial conditions}
We show that the current SCGF for the finite system with quenched initial condition reduces to that of the semi-infinite system in the limit $L\rightarrow\infty$. In the following, we denote $\rho_0$, $H_1$ for finite system as $\rho_0^{fi}$, $H_1^{fi}$. A straightforward calculation leads to
\begin{align}
  \lim\limits_{L\rightarrow\infty}\rho_0^{fi}(x,t)=\rho_0^{si}(x,t),\qquad \lim\limits_{L\rightarrow\infty} H_1^{fi}(x,t)=-1+\erf\qty(\frac{x}{\sqrt{4D(T-t)}})=H_1^{si}-1.
\end{align}
From above, it follows that $\partial_xH_1^{fi}(x,t)=\partial_xH_1^{si}(x,t)$ in the limit $L\rightarrow\infty$. Therefore,
\begin{align}
  \lim\limits_{L\rightarrow\infty}\int_0^Tdt\int_0^Ldx\, \sigma(\rho_0^{fi})(\partial_xH_1^{fi})^2=\int_0^Tdt\int_0^{\infty}dx\, \sigma(\rho_0^{si})(\partial_xH_1^{si})^2.
\end{align}
The result of finite-system is found to be consistent with the semi-infinite case as $L\rightarrow\infty$.

\subsubsection*{Annealed initial condition}
We show that the current SCGF for the finite system with annealed initial condition reduces to that of semi-infinite system in the limit $L\rightarrow\infty$. Since the first term in $\eqref{eq;generalanneal2}$ is identical to that for the quenched case, it suffices to check the limit of the second term.
\begin{align*}
  \lim\limits_{L\rightarrow\infty}\int_0^Ldx\,\frac{\sigma(\bar{\rho})}{2D}(H_1(x,0))^2&=\frac{\sigma{\brho}}{2D}\int_0^{\infty}dx\,\qty(-1+\erf\qty(x/\sqrt{4DT}))\\
  &=\frac{\sigma(\brho)\sqrt{T}}{\sqrt{D}}\int_0^{\infty}dx\,\qty(\erfc(x))^2\\
  &=\frac{(2-\sqrt{2})\sigma(\brho)\sqrt{T}}{\sqrt{D\pi}}
\end{align*}
The result of finite-system is found to be consistent with the semi-infinite case as $L\rightarrow\infty$.

\section{Current fluctuations for NESS}\label{Appendix_ness}
In this appendix, we derive $\eqref{eq:long_time_av}$; namely, we show that the long-time average of current variance for a finite system with a constant diffusion coefficient and an arbitrary mobility is consistent with the result for NESS. We need to evaluate
\begin{align}
  \lim\limits_{T\rightarrow\infty}\frac{\ev{Q_T^2}_{c,\mA}}{T}=\lim\limits_{T\rightarrow\infty}\frac{1}{T}\qty[\int_0^Tdt\int_0^Ldx \sigma(\rho_0)(\partial_xH_1)^2+\int_0^Ldx\frac{\sigma(\bar{\rho})}{2D}(H_1(x,0))^2].
\end{align}
The first term can be evaluated by considering the long-time relaxation of $\rho_0$ and $H_1$. As $T \rightarrow \infty$, the integrand relaxes smoothly to its steady-state value without any oscillation; the density profile approaches the NESS profile ($\rho_0(x,t) \rightarrow \rho^{\text{st}}(x)$), while the gradient of the auxiliary field settles to its stationary value ($\partial_x H_1 \rightarrow 1/L$) except near the final time $t \sim T$. Therefore, the long-time average is exclusively dominated by this steady-state value:
\begin{align*}
  \lim_{T\rightarrow\infty} \frac{1}{T} \int_0^T dt \int_0^L dx \, \sigma(\rho_0)(\partial_x H_1)^2 
  &= \lim_{t\rightarrow\infty} \int_0^L dx \, \sigma(\rho_0)(\partial_x H_1)^2 \nonumber \\
  &= \frac{1}{L^2} \int_0^L \sigma(\rho^{\text{st}}) \, dx.
\end{align*}
Meanwhile, the second term obviously vanishes in the long-time average,
\begin{align*}
  \lim_{T\rightarrow\infty} \frac{1}{T} \int_0^L dx \, \frac{\sigma(\overline{\rho})}{2D} (H_1(x,0))^2 = 0,
\end{align*}
since the spatial integral of $(H_1(x,0))^2$ remains finite as $T \rightarrow \infty$. Combining these results, we obtain
\begin{align}
   \lim\limits_{T\rightarrow\infty}\frac{\ev{Q_T^2}_{c,\mA}}{T}= \lim\limits_{T\rightarrow\infty}\frac{\ev{Q_T^2}_{c,\mQ}}{T}=\frac{1}{L^2}\int_0^L\sigma(\rho^{st})dx.
\end{align}

\section{Equivalence of RBM and IRW}\label{equivalence}
Under the diffusive scaling limit $\Lambda \to \infty$, both the discrete-lattice IRW and the continuous-space RBM can be mapped to the same macroscopic hydrodynamic description. By defining the density for IRW as the number of particles per site and for RBM as the number of particles per unit length $(1/\Lambda)$, both models exhibit $D(\rho)=1$, $\sigma(\rho)=2\rho$ (\cite{Dean1996,Derrida2009}) in the macroscopic limit $\Lambda\rightarrow\infty$. Furthermore, the boundary transition rates after scaling by $\Lambda$ are given by
\begin{table}[H]
  \centering
  \label{tab:my_label}
  \begin{tabular}{c|cc} 
                            & RBM  & IRW \\ \hline
    jump rate from reservoir & $A/\Lambda$ & $A/\Lambda$ \\ 
     jump rate to reservoir & $C\rho(0,t)/\Lambda$ & $Cn_1/\Lambda\simeq C\rho(0,t)/\Lambda$
  \end{tabular}
\end{table}
\noindent{}As shown in the table, the scaled boundary dynamics are identical for both RBM and IRW. Since the transport coefficients $D(\rho)$, $\sigma(\rho)$ and the boundary conditions are identical, the large deviation functions of the current for the RBM and IRW become identical within the MFT framework. Consequently, the governing MFT equations and their corresponding boundary conditions for the RBM reduce to those for the equivalent IRW system.

\section{Derivation of Current SCGF via Microscopic Calculation} \label{AppendixC}
In this appendix, we provide a microscopic derivation of the SCGF for the current in the model treated in Section $\ref{sectionRBM}$, specifically for the case of annealed initial condition. For annealed initial condition, the integrated current can be represented as a combination (sum and difference) of independent Poisson processes.
Let $Q_T^{L}$, $Q_T^{R}$ and $Q_T^{sys}$ be the integrated number of particles over the time interval $[0,T]$ satisfying the following conditions:
\begin{itemize}
  \item Net number of particles injected from reservoir L ; $Q_T^{L}$\\
  \item Number of particles injected from reservoir R and removed by reservoir L ; $Q_T^{R}$\\
  \item Number of particles, initially in the system, removed by reservoir L ; $Q_T^{sys}$\\
\end{itemize}
The integrated current $Q_T$ is written as $Q_T=Q_T^{L}-Q_T^{R}-Q_T^{sys}$. $Q_T^{L}$, $Q_T^{R}$ and $Q_T^{sys}$ follow Poisson distribution in the case of annealed initial condition.\\

\subsubsection*{The distribution of $Q_T^{L}$}
We define $P_L(x,t)$ as the probability density function of a particle at $t$ after being injected from reservoir L. The function $r_L(t)$ is the probability that a particle injected from Reservoir L is removed by Reservoir L within time interval $[0,t]$. The function $r_L(t)$ also satisfies $r_L(t)=-\int_0^t\pdv{P_L(x,t')}{x}dt'$.
$Q_T^{L}$ satisfies Poisson distribution with mean $A r_L(t)$. Thus, we have to find $P_L(x,t)$.

The distribution $P_L(x,t)$ satisfies the diffusion equation
\begin{align}
  \partial_tP_L(x,t)=\partial_x^2P_L(x,t), \quad (0<x<L).
\end{align}
Initial conditions and boundary conditions are 
\begin{align*}
  &P(x,0)=\delta(x-\varepsilon)\\
  &\partial_xP(x=0,t)=C P(0,t),\qquad \partial_xP(x=L,t)=-B P(L,t)
\end{align*}
By solving this and taking $\varepsilon\rightarrow0$,
\begin{align}
  P_L(x,t)=G^R(0,x;t)
\end{align}
$G^R(x,y;t)$ is the Green's function which is introduced in $\eqref{finite green fun}$.\\
From this, $r_L(t)$ is 
\begin{align}
 r_L(t)&=\int_0^{t}\pdv{P_L(x=0,t)}{x}dt=\int_0^t\partial_xG^R(0,x;\tau)d\tau
\end{align}
$Q_t^{L}$ follows this Poisson distribution,
\begin{align}
  P(Q_T^L=n)&=\frac{\Lambda_L^ne^{-\Lambda_L}}{n!}
\end{align}
with the parameter (average number of particles)
\begin{align}
  \Lambda_L&=\int_0^TA(1-r(T-t)) dt\notag \\
  &=A T-\int_0^Tdt\int_0^td\tau A G^R_x(0,x=0,\tau)\notag \\
  &=A T-\sum\limits_{n=1}^{\infty}\frac{AC(Tk_n^2+e^{-k_n^2T}-1)}{N_nk_n^4}
\end{align}

\subsubsection*{The distribution of $Q_T^{R}$ }
We define $P_R(x,t)$ as the probability distribution function of a particle at $t$ after being injected from reservoir R. The function $r_R(t)$ is the probability that a particle injected from Reservoir R is removed by Reservoir L within time interval $[0,t]$. The function $r_R(t)$ satisfies $r_R(t)=-\int_0^t\pdv{P_R(x,\tau)}{x}d\tau$.
Similar to the case of $Q_T^L$,
\begin{align}
  P_R(x,t)=G(L,x;t)
\end{align}
and $r_R(t)$ is
\begin{align}
  r_R(t)=\int_0^td\tau\partial_x G^R(L,x=0;\tau)d\tau
\end{align}
 $Q_T^{R}$ follows the Poisson distribution
\begin{align}
  P(Q_T^R=n)&=\frac{\Lambda_R^ne^{-\Lambda_R}}{n!}
\end{align}
with the parameter (average number of particles)
\begin{align}
  \Lambda_R&=\int_0^TDr_R(T-t)\notag \\
  &=\sum\limits_{n=1}^{\infty}\left(\cos k_nL+\frac{C}{k_n}\sin k_nL \right)\frac{(Tk_n^2+e^{-k_n^2T}-1)CD}{N_nk_n^4}
\end{align}

\subsubsection*{The distribution of $Q_T^{sys}$}
We define $P_{sys}(x,t|x')$ as the probability density function of a particle which is released from $x=x'$ at $t=0$. The function $r_{sys}(t)$ is the probability that a particle starts from $x=x'$ is removed by Reservoir L within time interval $[0,t]$. The function $r_{sys}(t)$ satisfies $r_L(t)=-\int_0^t\pdv{P_{sys}(x,t|x')}{x}dt'$.
The distribution $P_{sys}(x,t|x')$ satisfies this diffusion equation and its boundary conditions
\begin{align*}
  \partial_tP(x,t)&=\partial_x^2P(x,t),\qquad (0<x<L)\\
  \partial_xP(x=0,t)&=C P(0,t),\quad \partial_xP(L,t)=-B P(L,t),\quad P(x,0)=\delta(x-x')
\end{align*}
Solving these equations, we obtain
\begin{align*}
  P_{sys}(x,t|x')=G^R(x,x';t)
\end{align*}
Therefore, the expectation value of the number of particles ($\Lambda_{sys}$) passing the origin during time T is
\begin{align}
  \Lambda_{sys}&=\int_0^T\int_0^L\sum\limits_{n=1}^{\infty}\frac{C X_n(x')}{N_n}e^{-k_n^2}\bar{\rho}dx'dt \notag\\
  &=\sum\limits_{n=1}^{\infty}\frac{C\bar{\rho}}{N_n}\qty{\frac{\sin k_nL}{k_n}+\frac{C}{k_n^2}\{1-\cos k_nL\}}\qty{\frac{1-e^{-k_n^2T}}{k_n^2}}\notag\\
  &=\sum\limits_{n=1}^{\infty}\frac{C\bar{\rho}(k_n\sin k_nL+C(1-\cos k_nL))}{N_nk_n^4}\qty{1-e^{-k_n^2T}}
\end{align}
 $Q_T^{sys}$ follows Poisson distribution of average $\Lambda_{sys}$.

\subsubsection*{The CGF of $Q_T$}
Finally, we calculate the distribution of the integrated current $Q_T$. Because $Q_T^{R}+Q_T^{sys}$ follows a Poisson distribution with a mean of $\Lambda_R+\Lambda_{sys}$, $Q_T=Q_T^{L}-(Q_T^{R}+Q_T^{sys})$ follows the skellam distribution. Therefore, 
\begin{align*}
\mu(\lambda)=\Lambda_L(e^{\lambda}-1)+(\Lambda_R+\Lambda_{sys})(e^{-\lambda}-1)  
\end{align*}
From above results, current SCGF is 
\begin{align}\label{cumulant micro}
 \mu(\lambda) ={}& \biggl\{ A T - \sum_{n=1}^{\infty} \frac{AC(Tk_n^2+e^{-k_n^2T}-1)}{N_nk_n^4} \biggr\} (e^{\lambda}-1) \nonumber \\
 & + \biggl\{ \sum_{n=1}^{\infty} \left(\cos k_nL+\frac{C}{k_n}\sin k_nL \right) \frac{(Tk_n^2+e^{-k_n^2T}-1)C\delta}{N_nk_n^4} \nonumber \\
 & \qquad +\sum\limits_{n=1}^{\infty}\frac{C\bar{\rho}(k_n\sin k_nL+C(1-\cos k_nL))}{N_nk_n^4}\qty(1-e^{-k_n^2T})\biggr\} (e^{-\lambda}-1)
\end{align}
 where
\begin{align*}
  &N_n=\frac{L}{2}\left(1+\frac{C^2}{k_n^2} \right)+\frac{1}{4k_n}\left(1-\frac{C^2}{k_n^2} \right)\sin 2k_nL+\frac{C}{k_n^2}\sin^2k_nL\\
  &(B+C)k_n=(k_n^2-BC)\tan(k_nL)\label{eq;k_n}
\end{align*}
By using the relation $1-\sum\limits_{n=1}^{\infty}\frac{C}{N_nk_n^2}=\frac{1/C}{L+\frac{1}{B}+\frac{1}{C}}$\footnote{Setting $f(x)=-\frac{1}{L+\frac{1}{B}+\frac{1}{C}}x+\frac{L+\frac{1}{B}}{L+\frac{1}{B}+\frac{1}{C}}$ and expanding it with $X_n(x)$ from Eq. $\eqref{eq;Xn}$, we find $f(x)=\sum\limits_{n=1}^{\infty}\frac{C}{N_nk_n^2}X_n(x)$. Then, substituting $x=0$ leads to $\frac{L+\frac{1}{B}}{L+\frac{1}{B}+\frac{1}{C}}=\sum\limits_{n=1}^{\infty}\frac{C}{N_nk_n^2}$, which shows that $1-\sum\limits_{n=1}^{\infty}\frac{C}{N_nk_n^2}=\frac{1/C}{L+\frac{1}{B}+\frac{1}{C}}$.}, we can confirm that $\eqref{cumulant MFT}$ and $\eqref{cumulant micro}$ are identical.

\end{document}